\documentclass[12pt,a4paper]{article}
\usepackage[utf8]{inputenc}
\usepackage[T1]{fontenc}
\usepackage[english]{babel}
\usepackage[a4paper,top=2cm,bottom=2cm,left=2cm,right=2cm,marginparwidth=1.75cm]{geometry}
\usepackage{authblk}
\usepackage{balance}
\usepackage{bm}
\usepackage{siunitx}
\usepackage{graphicx}
\usepackage{array}
\usepackage[usenames,dvipsnames]{xcolor}
\usepackage{booktabs}
\usepackage{multirow}
\usepackage{epstopdf}
\usepackage{caption}
\usepackage{subcaption}
\usepackage[version=3]{mhchem}
\usepackage{soul}
\usepackage{url}
\usepackage{hyperref}
\usepackage{doi}
\usepackage{amsmath, amssymb}
\usepackage[export]{adjustbox}
\usepackage[numbers]{natbib}
\usepackage[prefix]{nomencl}
\usepackage{changes}
\usepackage{svg}
\usepackage{tikz}
\usepackage{float}
\usetikzlibrary{tikzmark}
\usepackage{pdflscape}
\usepackage{mathrsfs}
\usepackage{placeins}
\DeclareSIUnit\bar{bar}
\DeclareSIUnit\px{px}

\makenomenclature

\renewcommand{\nomgroup}[1]{%
  \ifthenelse{\equal{#1}{A}}{\item[\textbf{}]}{%
  \ifthenelse{\equal{#1}{B}}{\item[\textbf{Greek Symbols}]}{%
  \ifthenelse{\equal{#1}{C}}{\item[\textbf{Subscripts}]}{}}}}
\hypersetup{colorlinks = true,linkcolor = orange,anchorcolor =red,citecolor = black,filecolor = red,urlcolor = red, pdfauthor=author}

\title{Bubble-induced versus thermodynamic voltage losses during pressurized alkaline water electrolysis}

\author[1,2]{Hannes Rox\footnote{Corresponding authors: h.rox@hzdr.de,  k.eckert@hzdr.de}} 
\author[3,4]{Feng Liang}
\author[5]{Robert Baumann}
\author[6]{Mateusz M. Marzec}
\author[6]{Krystian Sokołowski}
\author[1]{Xuegeng Yang}
\author[5,7]{Andrés F. Lasagni}
\author[3]{Roel van de Krol} 
\author[1,2,8]{Kerstin Eckert$^*$} 

\affil[1]{Institute of Fluid Dynamics, Helmholtz-Zentrum Dresden-Rossendorf, 01328 Dresden, Germany. E-mail: h.rox@hzdr.de; k.eckert@hzdr.de}
\affil[2]{Institute of Process Engineering and Environmental Technology, Technische Universität Dresden, 01062 Dresden, Germany.}
\affil[3]{Institute for Solar Fuels, Helmholtz-Zentrum für Materialien und Energie GmbH, 14109 Berlin, Germany.}
\affil[4]{School of Mechanical Engineering, Xi’an Jiaotong University, Xi'an, Shaanxi 710049, China.}
\affil[5]{Institute of Manufacturing, TU Dresden, 01062 Dresden, Germany.}
\affil[6]{Academic Centre for Materials and Nanotechnology, AGH University of Krakow, 30-059 Krakow, Poland.}
\affil[7]{Fraunhofer Institute for Material and Beam Technology IWS, Winterbergstraße 28, 01277 Dresden, Germany.}
\affil[8]{Hydrogen Lab, School of Engineering, Technische Universität Dresden, 01062 Dresden, Germany.}

\begin{document}
\date{}	
\maketitle

\begin{abstract}
Understanding how bubbles influence the efficiency of water electrolysis is crucial to achieve economically competitive hydrogen, generated by renewable energy sources, such as wind and solar power. Water electrolysis is typically performed at high pressures to reduce the cost of energy-intensive mechanical compression of the produced \ce{H2}. Thus, a better understanding of how the absolute pressure affects electrochemical performance and bubble size is necessary. In general, bubble sizes decrease as the pressure increases. Using different-sized pillar-patterned Ni electrodes generated by Direct Laser Writing, the detached bubble sizes can be modified even at elevated pressures. As the pillar size increases, the bubbles become larger at all pressures investigated from 1 to \SI{6}{\bar}. At a current density of \SI{-25}{\milli\ampere\per\centi\metre\squared}, the cathodic potential increases with pressure according to the thermodynamic voltage losses given by the Nernst equation ($\approx$ \SI{23}{\milli\volt} at $p=\SI{6}{\bar}$). Surprisingly, increasing the current density to \SI{100}{\milli\ampere\per\centi\metre\squared} leads to a reduction of the overpotential by up to $\approx$ \SI{60}{\milli\volt}. Reduced bubble sizes at increased pressures minimize the losses caused by the bubbles, thereby compensating for the thermodynamic voltage penalty. Applying the Buckingham $\Pi$-theorem enables the derivation of dimensionless numbers to characterize the ratio of bubble-induced and thermodynamic voltage losses.
\end{abstract}

\section*{Broader Context} 
The Belém Declaration, recently launched at COP30, strengthens the global agenda for green industrialization to accelerate the decarbonization of heavy emitting industries. Therefore, a rapid increase in the use of low-emission technologies is needed to reduce greenhouse gas emissions by the end of the decade. One of these low-emission technologies is green hydrogen, produced by water electrolysis using renewable energy sources like solar or wind power. 
However, although green hydrogen reduces environmental impact, the costs of water electrolysis remains high, at up to eight times that of gray hydrogen. Therefore, reducing the operating and capital expenditures by improved electrode and cell designs is essential.
Considerable losses are caused by the evolving hydrogen and oxygen bubbles by increasing ohmic resistances and blocking electrochemically active sites. Consequently, a better understanding of the effects of electrogenerated bubbles on the electrochemical reaction is needed to improve the overall efficiency and reduce the costs of green hydrogen production.
\section{Introduction} 
\label{sec:intro}
Understanding the ongoing processes of gas evolution on electrodes is essential for many electrochemical systems, including water splitting \cite{Zhang2012}, chloro-alkali electrolysis \cite{Botte2014} or \ce{CO2} electroreduction \cite{Lim2014}. Over the past two decades, many studies have focused on the hydrogen evolution reaction (HER) and the oxygen evolution reaction (OER) because they are essential to improve the overall efficiency of water electrolysis \cite{Emam2024, Mahmood2024, Yang2020}. This is necessary to replace fossil fuels with hydrogen produced using low- or zero-carbon energy sources, such as solar or wind-derived electricity \cite{Hermesmann2021}. Nevertheless, the efficiency and economic competitiveness of large-scale green hydrogen production are still lacking.

The presence of evolving hydrogen and oxygen bubbles can lead to significant losses due to increasing ohmic resistance and blocking of electrochemically active sites \cite{He2023, Swiegers2021, Angulo2020}. Additionally, the mass transfer and the actual current density are influenced by electrode coverage \cite{Vogt2017, Vogt2015, Vogt2011, Eigeldinger2000}. Various studies have shown that mitigating these bubble effects can significantly lower the electrode potentials by 9-\SI{25}{\%} \cite{Li2009, Wang2010, Tiwari2019}. The reduction of the bubble-induced losses can be achieved by optimizing the electrode's electrocatalysts \cite{Jiao2021}, morphology \cite{Rocha2022, Fujimura2021, Darband2019} or surface \cite{Darband2019,Li2023,Andaveh2022}, enhancing bubble coalescence \cite{Bashkatov2024, Lv2021}, or applying external forces \cite{Rocha2022,  Darband2019, Li2021b,Baczyzmalski2017, Koza2011}.  However, process parameters like the absolute pressure in the cell also influence the bubble evolution. Since industrial electrolyzers are usually operated under elevated pressures up to 30-\SI{50}{\bar} \cite{Emam2024}, the understanding of the influence of the pressure on the gas evolution during water splitting reaction is needed. 

To the best of the authors knowledge, the only comprehensive study reporting the characterization of electrogenerated bubbles at higher pressures in alkaline electrolyte (\SI{1}{M} KOH) is the work of \citeauthor{Sillen1983} \cite{Sillen1983}. Here, both oxygen and hydrogen bubbles generated under various pressures ($p = 0.25-\SI{30}{\bar}$) were investigated. After this early work, \citeauthor{Liang2024} \cite{Liang2024} and \citeauthor{Luo2024} \cite{Luo2024} studied the impact of elevated pressure on photoelectrochemical water splitting. In general, the increase in pressure leads to an increase of the dissolved \ce{H2} concentration \cite{Liang2024} and decrease in bubble size \cite{Sillen1983,Liang2024, Luo2024}. Additionally, the thermal behavior of the electrolyzer is significantly impacted by pressure. Increasing pressure leads to reduced water vaporization and thus reduced water losses \cite{Ogata1988, LeRoy1983}. Furthermore, the rising pressure also increases the thermodynamic voltage losses \cite{Liang2024, Ogata1988, LeRoy1983, Rauschenbach2015}. This can be understood by reformulating the Nernst equation in terms of the absolute pressure $p$ \cite{Pfennig2025,Espinosa2018}. Therefore, the dependence of the potential $E$ on the partial pressure $p_\mathrm{\ce{H2}}$ and $p_\mathrm{\ce{O2}}$ can be defined as 
\begin{align}
    E = E^0 + \frac{\mathscr{R}T}{zF}\ln \left(\frac{p_\mathrm{\ce{H2}}}{p_0} \sqrt{\frac{p_\mathrm{\ce{O2}}}{p_0}}\right), \label{eq:nernst}
\end{align}
where $E^0$ denotes the standard potential, $\mathscr{R}$ the universal gas constant, $z$ the charge number, $F$ the Faradaic constant and $p_0$ the atmospheric pressure. In Eq.~\ref{eq:nernst}, it is assumed that the fugacity coefficients of \ce{H2} and \ce{O2} are equal to 1. This is a reasonable assumption since, at a pressure of \SI{10}{\bar}, the deviation from the Nernst potential using the actual fugacity coefficients for \ce{H2} (1.03) and for \ce{O2} (0.92) is only \SI{-0.5}{\milli\volt} \cite{Vallejos-Burgos2019}.

In the present case of a symmetrical water electrolyzer with a constant pressure at both anode and cathode due to the absence of a separator, it can be assumed that $p_\mathrm{\ce{H2}} = p_\mathrm{\ce{O2}} = p$ \cite{Millet2022}. Here, $p$ denotes the absolute pressure of the electrolyzer. Furthermore, by neglecting the saturating pressure of water vapor in \ce{H2} and \ce{O2}, Eq.~\ref{eq:nernst} can be reformulated as
\begin{align}
    E = E^0 + \frac{\mathscr{R}T}{zF}\ln \left(\left(\frac{p}{p_0}\right)^{3/2}\right).
\end{align}
Assuming a constant ambient temperature $T=\SI{298.15}{\kelvin}$ and pressure $p_0 = \SI{1}{\bar}$, and thus a constant standard potential $E^0$, the theoretical pressure-induced thermodynamic losses $E_\mathrm{theor}$ can be calculated with the second term of the above equation \cite{Millet2022}:
\begin{align}
    E_\mathrm{theor} = \frac{\mathscr{R}T}{zF}\ln \left(\left(\frac{p}{p_0}\right)^{3/2}\right) \label{eq:theory_td_losses}
\end{align}
This leads to additional cell voltages of $\approx$ \SI{66}{\milli\volt} when increasing the operating pressure from \SI{1}{\bar} to \SI{30}{\bar}.

In contrast, bubble-induced losses are lowered as the bubble sizes are decreased simultaneously \cite{Sillen1983, Liang2024}. However, this effect has not been thoroughly investigated for alkaline water electrolysis, which is still the most mature technology to produce green hydrogen and is typically performed at current densities of 100 to \SI{500}{\milli\ampere\per\centi\metre\squared} \cite{ElEmam2019}. \citeauthor{Sillen1983} focused primarily on the evolving bubbles and did not discuss the influence of the operating pressure on the overpotential \cite{Sillen1983}. In addition, photoelectrochemical water splitting runs with a rather small current density of \SI{10}{\milli\ampere\per\centi\metre\squared} \cite{Liang2024}. The most recent work is from \citeauthor{Wu2025} \cite{Wu2025}, investigating the growth of a single hydrogen bubble on a Pt electrode with an aerophilic PTFE spot as nucleation site. In contrast to all previous work, larger bubbles are reported at elevated pressures up to \SI{20}{\bar}. \citeauthor{Wu2025} relate this to an increase in the aerophilicity of electrode surface  \cite{Wu2025}. However, as \ce{H2SO4} was used as electrolyte and only single bubbles are studied, these results are difficult to transfer to industrial alkaline electrolysis. Consequently, little is known about the influence of absolute pressure on bubble-induced losses in alkaline water electrolysis.

To address this gap, different bubble sizes are needed to see how they affect the overall electrode potential at elevated pressures. In \citeauthor{Rox2025b} we showed that the detached bubble size is significantly changed by pillar-like surface patterns produced via Direct Laser Writing (DLW) \cite{Rox2025b}. This is due to the dual wetting behavior of these surfaces, which can be understood as a hierarchical surface with defined hydrophobic nucleation sites incorporated into a system of hydrophilic microchannels. Based on the current understanding of the bubble evolution on the pillar-patterned electrodes, the bubbles are pinned to the hydrophobic patches during their growth \cite{Rox2025b}. In addition, the forces acting on the growing bubbles \cite{Bashkatov2022, Hossain2022} are modified by these surface patterns. Here, especially the interfacial tension force $F_\mathrm{S}$ (see. Eq.~\ref{eq:interfacial_tension_force}) and the contact pressure force $F_\mathrm{CP}$ (see. Eq.~\ref{eq:contact_pressure_force}) are influenced by the modified contact angle $\theta$ and contact radius $r_\mathrm{c}$ of the bubble:
\begin{align}
        F_\mathrm{S} &= -2 \cdot \pi \cdot r_\mathrm{c} \cdot \gamma \cdot \sin \theta \cdot f_\mathrm{S} \label{eq:interfacial_tension_force} \\
        F_\mathrm{CP} &= \frac{2\cdot \gamma}{r_\mathrm{B}} \cdot \pi \cdot r_\mathrm{c}^2 \label{eq:contact_pressure_force}
\end{align}
A schematic drawing of these two forces acting on a bubble growing at a horizontal electrode can be found in the ESI (see Fig.~\ref{fig:appendix_bubble_forces}). $F_\mathrm{S}$ acts as a retarding force at the three-phase contact line with radius $r_\mathrm{c}$, drawing the bubble with radius $r_\mathrm{B}$ towards the electrode surface due to the surface tension $\gamma$ \cite{Bashkatov2022, Demirkır2024, Meulenbroek2021}. On rough surfaces, the three-phase contact line can be cut into a discontinuous state by electrolyte-filled pockets below the bubble, which can be taken into account by the surface area fraction $f_S$ \cite{Li2023}. In contrast, $F_\mathrm{CP}$ acts as a detaching force \cite{Bashkatov2022}, which dominates when $\sin \theta$ approaches 0. Consequently, changing the size of the hydrophobic nucleation sizes results in a change of $r_\mathrm{c}$ and therefore is expected to change the size of the detached bubbles.

In the present study, different sized pillar-patterns are applied to high-purity Ni-foils using DLW to modify the detached bubble size. This enables the study of the influence of detached bubble size and absolute cell pressure on the cathodic potential during HER in an alkaline electrolyte at elevated current densities and pressures. Linking these results with the theoretical principles of alkaline water electrolysis provides new insights into the importance of minimizing bubble-induced losses to improve the overall efficiency.

\section{Results and discussion}
\label{sec:results}

\subsection*{Electrode characterization}
\begin{figure}[ht]
    \centering
    \includegraphics[width=\linewidth]{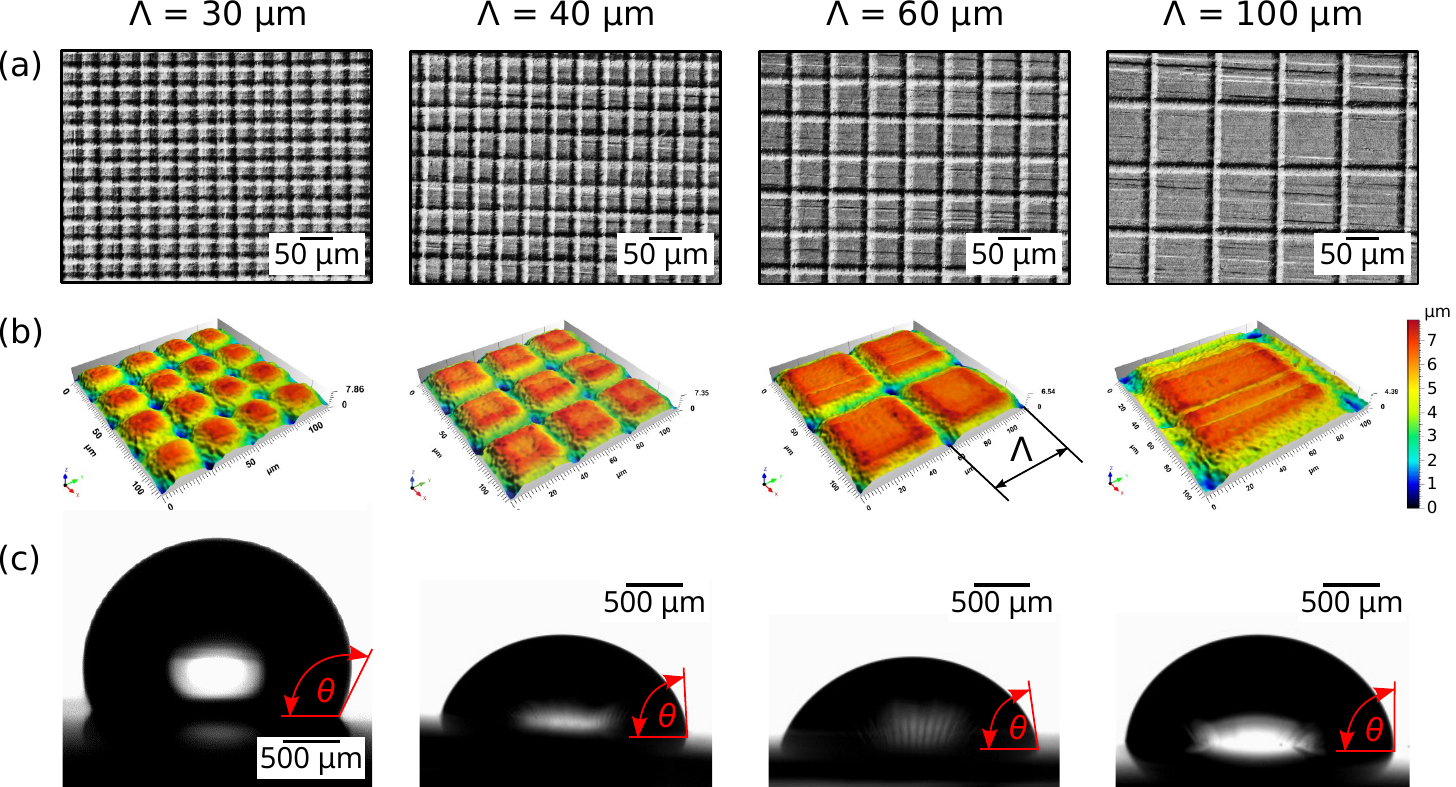}
    \caption{Overview of all laser-structured electrodes: (a) Microscopic and  (b) confocal images of electrode structures with definition of the spatial period $\Lambda$. (c) Wetting behavior of electrode surfaces with applied water droplet and highlighted contact angles $\theta$}.
    \label{fig:pressure_electrodes}
\end{figure}

To ensure maximum flexibility in terms of structure size and thus, detached bubble size, DLW was used to texture the electrode surfaces. The side length of the pillars, here referred to as spatial period $\Lambda$, was changed from \SI{30}{\micro\metre} to \SI{100}{\micro\metre} at a constant depth of $\approx$ \SI{7}{\micro\metre}.  A total of 4 different laser structures were studied, as shown in Fig.~\ref{fig:pressure_electrodes}. In addition, one non-structured electrode (NSE) was used as comparison. A summary of all characteristics of the studied electrodes is provided in Table~\ref{tbl:p_electrode_metrics}. 

\begin{table}[!b]
    \centering
    \caption{Spatial period of pillar structure $\Lambda$, average surface roughness $S_\mathrm{a}$, developed interfacial area ratio $S_\mathrm{dr}$, and water contact angle $\theta$ of all electrodes in comparison to the non-structured Ni-foil (NSE).}
    \label{tbl:p_electrode_metrics}
    \begin{tabular*}{0.6\textwidth}{@{\extracolsep{\fill}}llll}
    \hline
    \bm{$\Lambda$} & \bm{$S_\mathrm{a}$} & \bm{$S_\mathrm{dr}$}&\bm{$\theta$} \\
     (\si{\micro\metre}) & (\si{\micro\metre}) & (\%)& (\si{\degree})\\
    \hline
    \multicolumn{3}{c}{------ NSE ------}  & $55.2 \pm 0.5$  \\
    30  & 1.26 & 9.12 & $114.7 \pm 3.4$  \\
    40  & 1.16 & 6.70& $81.3 \pm 3.7$  \\
    60 & 0.927 & 3.64 & $77.0 \pm 3.2$  \\ 
    100  & 0.627 & 1.49 & $86.7 \pm 1.5$ \\
    \hline
  \end{tabular*}
\end{table}

Earlier works show that the average electrode roughness $S_\mathrm{a}$ and developed interfacial area ratio $S_\mathrm{dr}$ reflect well the change in electrochemically active electrode area \cite{Rox2025b, Rox2025a}. Here, $S_\mathrm{dr}$ is understood as the percentage of the additional surface area contributed by the texture compared to the planar definition area. Thus, the electrochemically active electrode area decreases with increasing spatial period $\Lambda$ (cf. Table~\ref{tbl:p_electrode_metrics}).

Based on Eq.~\ref{eq:interfacial_tension_force} and Eq.~\ref{eq:contact_pressure_force}, the contact angle $\theta$ of the hydrophobic pillars that facilitate the initial bubble nucleation is decisive for the detached bubble size. Therefore, the pillar-patterned electrodes were heated up for \SI{1}{\hour} at $\approx$ \SI{100}{\celsius} to enhance the hydrophobicity. This process can be understood as an accelerated version of the commonly reported aging  effect on laser-structured metal surfaces. During the heat treatment carbon compounds from the ambient air are adsorbed at the surface resulting in an increase of carbon content \cite{Heinrich2024,Kuisat2021,Schnell2020, Murillo2019,Yang2016}. This aging effect depends on the oxide layer established during the laser-structuring \cite{Heinrich2024}. Finally, in comparison to the previous study of \citeauthor{Rox2025b} \cite{Rox2025b} a more hydrophobic surface is achieved on top of the pillars. This was done to ensure that the hypothesized change in bubble size for different sized pillars holds true by strengthening the interfacial tension force $F_\mathrm{S}$ (see Eq.~\ref{eq:interfacial_tension_force}).

For the quantification of the change in surface chemistry, XPS was used. Details about the deconvolution and fitting of the high-resolution XPS spectra can be found in Sec.~\ref{sec:appendix_xps} and Fig.~\ref{fig:xps_spectra_c}-\ref{fig:xps_spectra_fe} in the supplementary information. For comparison, XPS analysis was also performed on a non-heat-treated electrode of each surface structure  (see Table~\ref{tbl:pressure_xps}). Here, it is shown that all surfaces of the heat treated electrodes are featuring a significant change in surface chemistry. As expected, the proportion of carbon compounds, like \ce{C-C} or \ce{C-H}, increased by $\approx$ \SI{10}{\%} for the laser-structured electrodes. In contrast, all \ce{Ni} compounds are drastically reduced. Here, the electrode with $\Lambda = \SI{30}{\micro\metre}$ exhibits the most significant change in surface composition in terms of its Ni$^0$ content, as it reaches 0. With increasing spatial period of the laser-structure even Fe$^{3+}$ was adsorbed at the surface, which was reported to improve HER, but also can result in electrode poisoning \cite{Demnitz2024}. In this work, however, we only consider the changes in wettability associated with the altered surface chemistry.

\begin{table}[ht]
\small
  \caption{Averaged surface composition (in atomic \%) of the electrodes determined by fitting high-resolution XPS spectra. Annotations: $^\star$Fresh electrode samples without heat treatment.}
  \label{tbl:pressure_xps}
  \begin{tabular*}{\textwidth}{@{\extracolsep{\fill}}lccccccccc}
    \toprule
    \textbf{Element} &\multicolumn{3}{c}{\textbf{C}} & \multicolumn{3}{c}{\textbf{O}} & \multicolumn{2}{c}{\textbf{Ni}} & \textbf{Fe}\\
    BE (eV) & 285.0 & 286.6 & 288.1 & 529.8 & 531.6 & 534.1 & 852.7 & 854.3 & 709.6 \\
    & C-C & C-OH & C=O & O-Ni & O-Ni, O-Fe, & O-C, & \multirow{3}{*}{Ni$^0$} & Ni$^{2+}$ & \multirow{3}{*}{Fe$^{3+}$} \\
    Groups/ & C-H & C-O-C & N-C=O & O-Fe & O=C, & -OH & & NiO &\\
    Ox. state & & C-N & & O-Cr & O-Si & & & &\\
    \midrule
    NSE$^\star$ & 32.1 & 12.2 & 3.3 & 10.0 & 14.6 & 8.5& 3.1 & 13.3 & 0.0\\
    NSE & 36.8 & 12.1 & 5.1 & 6.0 & 17.5 & 7.2 & 1.7 & 6.2 & 0.0 \\
    $\Lambda = \SI{30}{\micro\metre}^\star$& 23.8 & 7.3 & 3.4 & 13.6 & 18.7 & 5.9 & 3.1 & 20.4 & 0.0\\
    $\Lambda = \SI{30}{\micro\metre}$ & 37.0 & 16.3 & 5.4 & 1.8 & 20.0 & 8.5 & 0.0 & 6.1 & 0.0\\
    $\Lambda = \SI{40}{\micro\metre}^\star$ & 16.8 & 7.3 & 3.9 & 15.9 & 17.7 & 6.6 & 3.6 & 25.5 & 0.0\\
    $\Lambda = \SI{40}{\micro\metre}$ & 28.1 & 9.9 & 5.2 & 11.9 & 18.8 & 8.4 & 0.5 & 9.0 & 2.7 \\
    $\Lambda = \SI{60}{\micro\metre}^\star$ & 21.2 & 6.5 & 3.7 & 14.5 & 18.3 & 5.7 & 3.6 & 23.7 & 0.0\\
    $\Lambda = \SI{60}{\micro\metre}$ & 31.0 & 9.5 & 4.9 & 10.5 & 18.0 & 6.6 & 0.9 & 11.2 & 3.0 \\
    $\Lambda = \SI{100}{\micro\metre}^\star$ & 17.9 & 4.8 & 2.7 & 16.0 & 18.7 & 5.1 & 4.3 & 28.5 & 0.0 \\
    $\Lambda = \SI{100}{\micro\metre}$ & 27.3 & 8.3 & 4.6 & 10.8 & 21.6 & 6.7 & 0.5 & 9.5 & 4.3\\
    \bottomrule
  \end{tabular*}
\end{table}

To confirm the hydrophobicity of the electrode surfaces the contact angle $\theta$ of a \SI{2}{\micro\litre} water droplet was measured after the electrochemical measurements, as shown in Fig.~\ref{fig:pressure_electrodes} (c). For this purpose, five droplets were randomly placed on each surface, and the average contact angle was calculated using the arithmetic mean value. Here, the smallest electrode structure with a spatial period of $\Lambda = \SI{30}{\micro\metre}$ showed the largest contact angle $\theta$ of $\approx$ \SI{114.7}{\degree}. This electrode also showed the largest change in surface chemistry (cf. Table~\ref{tbl:pressure_xps}). The contact angles of all other laser-structured electrodes ($\Lambda \ge \SI{40}{\micro\metre}$) feature smaller values between \SI{77.0}{\degree} and \SI{86.7}{\degree}. According to the wetting theory, these surfaces are not hydrophobic as $\theta < \SI{90}{\degree}$ \cite{Butt2018}. However, it should be noted that the contact area of the applied droplet is much larger than the size of the laser-structured pillars. For example, the diameter of the contact area of the applied droplet at $\Lambda = \SI{30}{\micro\metre}$ is \mbox{$\approx$ 80} times larger than $\Lambda$. Thus, the measured contact angle $\theta$ can be understood as an average contact angle across $\approx$ 80 pillars and grooves, which presumably contain air pockets due to Cassie-Baxter wetting \cite{Bormashenko2017}. For the largest surface structure with $\Lambda = \SI{100}{\micro\metre}$, this reduces to \mbox{$\approx$ 30} pillars and grooves. In summary, the heat treatment strongly oxidized the laser-structured surfaces, achieving a more hydrophobic surface than the NSE surface, which exhibited a contact angle $\theta$ of \SI{55.2}{\degree}.

\subsection*{Influence of electrode structure size on detached bubble sizes at 1 bar}
\begin{table}[!b]
    \centering
    \caption{Overview of all studied experimental parameters.}
    \label{tbl:p_parameters}
    \begin{tabular*}{0.65\textwidth}{@{\extracolsep{\fill}}ll}
    \hline
    \textbf{Parameter} & \textbf{Range} \\
    \hline
    Spatial period $\Lambda$ (\si{\micro\metre}) & 0 (NSE), 30, 40, 60, 100 \\
    Current density $j$ (\si{\milli\ampere\per\centi\metre\squared}) & -25, -50, -100 \\ 
    Absolute pressure $p$ (\si{\bar}) & 1, 2, 4, 6 \\
    \hline
  \end{tabular*}
\end{table}
\begin{figure}[ht]
    \centering
    \includegraphics[width=\linewidth]{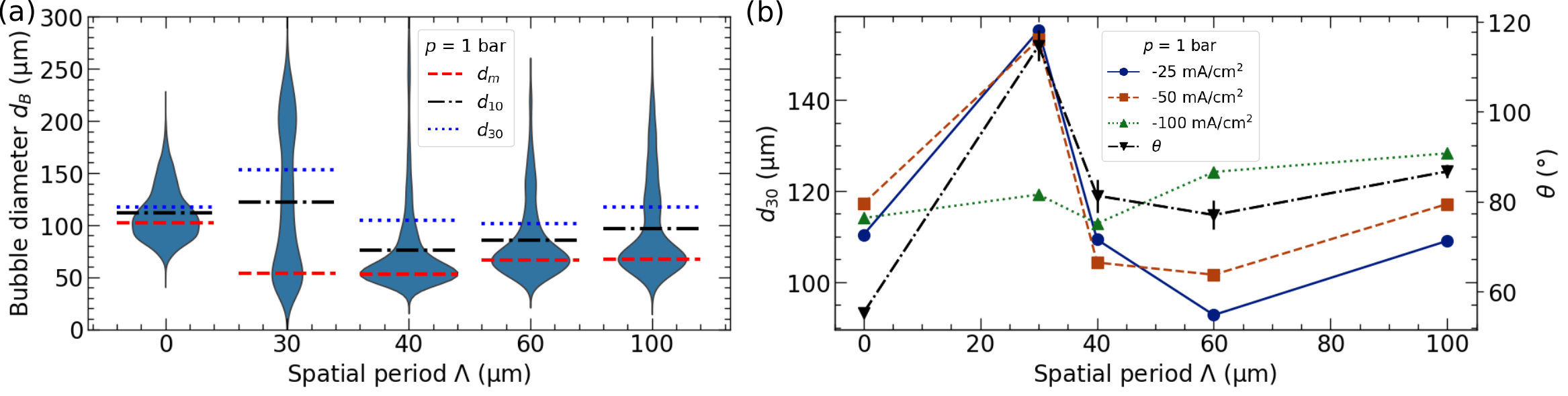}
    \caption{(a) Bubble size distribution at \SI{1}{\bar} and \SI{-50}{\milli\ampere\per\centi\metre\squared} depending on the spatial period $\Lambda$, where $\Lambda = \SI{0}{\micro\metre}$ corresponds to the non-structured electrode (NSE). (b) Volumetric mean diameter $d_{30}$ as a function of the applied current density $j$ and spatial period $\Lambda$ as well as the variation of the contact angle $\theta$ depending on spatial period $\Lambda$. The lines serve only to guide the reader's eye.}
    \label{fig:bubbles_1bar}
\end{figure}

The electrodes were then assembled vertically in a membrane-free, pressurized alkaline electrolyzer (see Fig.~\ref{fig:appendix_pressure_setup} in the supplementary information) to perform galvanostatic measurements in \SI{1}{M} KOH while simultaneously recording the evolving bubbles. The design of the cell was an adapted version of the setup used in \citeauthor{Liang2024} \cite{Liang2024}. Besides the electrode surface, both the absolute current density $j$ and the absolute pressure $p$ have been varied until \SI{100}{\milli\ampere\per\centi\metre\squared} and \SI{6}{\bar}, respectively, as detailed in Table~\ref{tbl:p_parameters}. The evolving bubbles captured on high-speed images were analyzed using Python 3.9 and StarDist (v.0.8.5) \cite{Schmidt2018, Weigert2020}. The image processing followed the segmentation method introduced in \citeauthor{Rox2023} \cite{Rox2023}.

To check the influence of surface modification on the bubble evolution, we first performed measurements at \SI{1}{\bar} and examined the detached bubble sizes as a function of the spatial period $\Lambda$ and the applied current density $j$. The shown bubble size distributions in Fig.~\ref{fig:bubbles_1bar} (a) with highlighted arithmetic mean diameter $d_\mathrm{10}$, mode value $d_\mathrm{m}$ and volumetric mean diameter $d_{30}$, reveal a non-trivial effect of $\Lambda$. Most laser-structured electrodes show a broader size distribution compared to NSE ($\Lambda = \SI{0}{\micro\metre}$). However, with the exception of the electrode with $\Lambda = \SI{30}{\micro\metre}$, the expected effect of larger detached bubbles by an increase in $\Lambda$ is well reflected. Due to the more hydrophobic electrode surface for $\Lambda = \SI{30}{\micro\metre}$ (cf. Fig.~\ref{fig:pressure_electrodes} (c) and Fig.~\ref{fig:bubbles_1bar} (b)) and thus, greater impact of the interfacial tension force $F_\mathrm{S}$, even larger bubbles are measured for this surface compared to $\Lambda = \SI{100}{\micro\metre}$. 

For better comparison of the results in the following the focus is set on the volumetric mean diameter $d_{30}$, which is defined as
\begin{align}
    d_{30} = \left(\frac{\sum_i n_\mathrm{i}d_\mathrm{B,i}^3}{N_\mathrm{B,tot}}\right)^{1/3},
\end{align}
where $n_\mathrm{i}$ is the number of bubbles with the size $d_\mathrm{B,i}$ and $N_\mathrm{B,tot}$ is total number of bubbles. The volumetric mean diameter, $d_{30}$, was chosen because it most accurately represents the total gas volume of all the mean values, making it a reliable indicator of the effect of pressure on the gaseous phase. In contrast to the results reported in \citeauthor{Rox2025b} \cite{Rox2025b}, the shown volumetric mean diameters in \mbox{Fig.~\ref{fig:bubbles_1bar} (b)} tend to increase with rising current densities and are in general $\approx$ 2 times larger compared to the previously reported data. This again shows that the heat treatment significantly changes the electrode surface properties and thus, the bubble dynamics. However, direct comparison is difficult since the electrodes were mounted horizontally in \cite{Rox2025b}. In general, the black curve in Fig.~\ref{fig:bubbles_1bar} (b) shows that the contact angle $\theta$ is decisive for the detached bubble size due to the two forces ($F_\mathrm{S}$ and $F_\mathrm{CP}$) mentioned above (see Eq.~\ref{eq:interfacial_tension_force} and \ref{eq:contact_pressure_force}). 

\subsection*{Influence of pressure on detached bubble sizes}

\begin{figure}[ht]
    \centering
    \includegraphics[width=\linewidth]{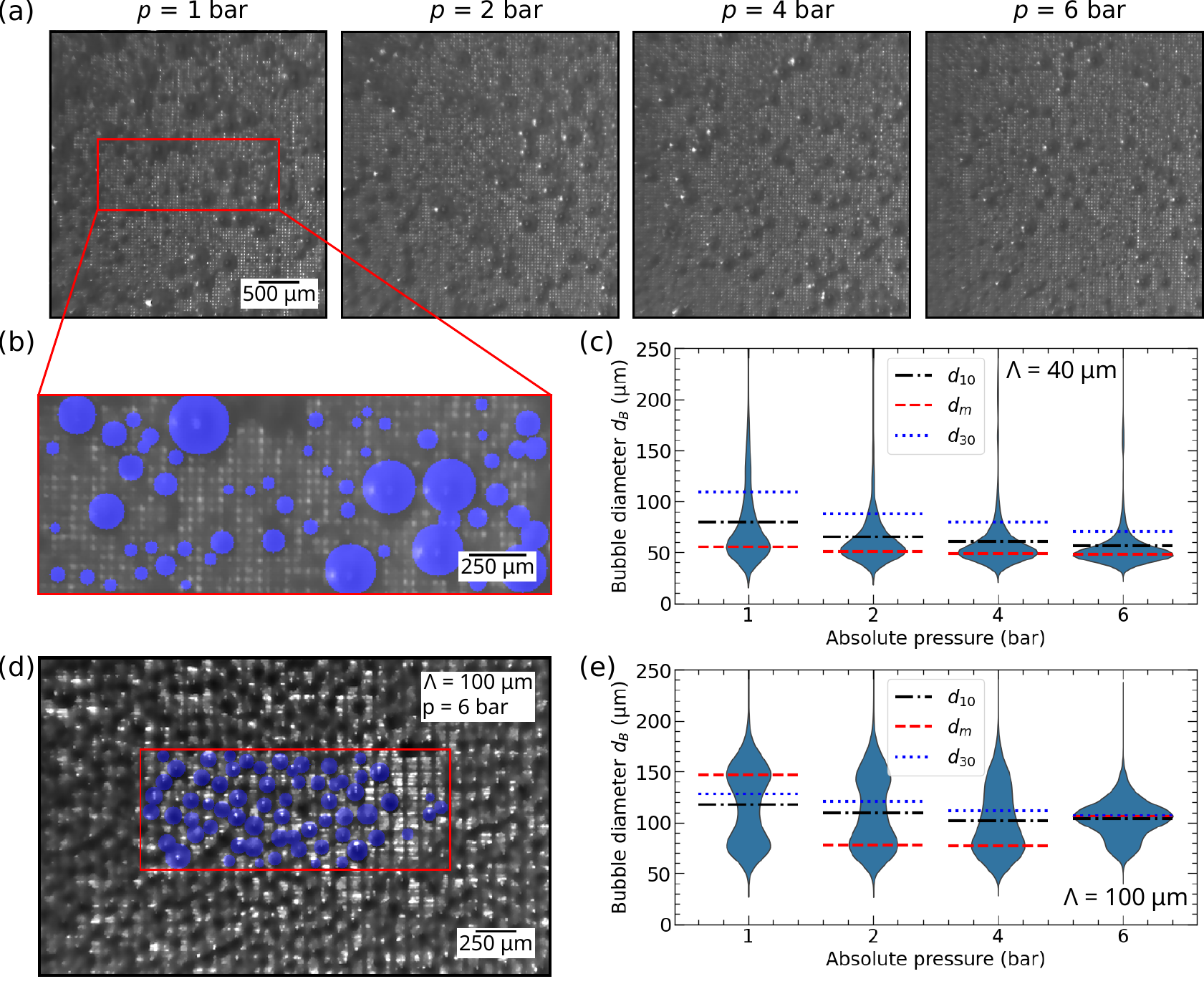}
    \caption{(a) Snapshots of the bubbles growing at the electrode with $\Lambda = \SI{40}{\micro\metre}$ at \SI{-100}{\milli\ampere\per\centi\metre\squared} depending on the absolute cell pressure $p$. (b) Example of the bubble segmentation using StarDist and (c) resulting bubble size distributions at \SI{-100}{\milli\ampere\per\centi\metre\squared}. (d) Nearly monodisperse bubble carpet at the electrode with $\Lambda = \SI{100}{\micro\metre}$ at \SI{-100}{\milli\ampere\per\centi\metre\squared} and \SI{6}{\bar} with highlighted bubble segmentations inside the region of interest. (e) Shift of bimodal to a nearly monodisperse bubble size distribution with increasing pressure for the electrode with $\Lambda = \SI{100}{\micro\metre}$ at \SI{-100}{\milli\ampere\per\centi\metre\squared}.}
    \label{fig:pressure_bubbles_snapshots}
\end{figure}

We will now examine the influence of absolute pressure on the detached bubble sizes. Example snapshots of the bubbles at the electrode with $\Lambda = \SI{40}{\micro\metre}$ at the highest current density ($j = \SI{-100}{\milli\ampere\per\centi\metre\squared}$) for all studied pressures are shown in Fig.~\ref{fig:pressure_bubbles_snapshots} (a). To improve the bubble tracking and reduce computational time, a rectangular area  measuring \SI{150}{\px} $\times$ \SI{400}{\px} was cropped out of the middle of each image and analyzed using StarDist (see  Fig.~\ref{fig:pressure_bubbles_snapshots} (b)). Hereby, the field of view was located in the center of the vertically mounted electrode to avoid artifacts from its cut edges. On the one hand, these edges exhibited different wetting behavior due to the silicone used (101RF, Microset, UK). On the other hand, the effective current density is greater than that of the bulk material due to the concentration of electric field lines \cite{Schoppmann2025}. Due to the upward movement of \ce{H2} bubbles caused by buoyancy, some of the bubbles captured in the images had detached from lower electrode areas.

In line with the findings of \citeauthor{Sillen1983} \cite{Sillen1983} and \citeauthor{Liang2024} \cite{Liang2024} in average the bubble sizes are decreased with increasing pressure. The shifting towards smaller bubble sizes with increasing pressure is well reflected in the calculated bubble size distribution in Fig.~\ref{fig:pressure_bubbles_snapshots} (c). In addition, the increase in absolute pressure led to a narrower bubble size distribution for all electrodes studied. As a result, the calculated mean diameters are closest to each other at highest pressure of \SI{6}{\bar}. This is particularly pronounced for the electrode with $\Lambda = \SI{100}{\micro\metre}$ at $j=\SI{-100}{\milli\ampere\per\centi\metre\squared}$. As shown in Fig.~\ref{fig:pressure_bubbles_snapshots} (e), all calculated mean values coincide. This results in a nearly monodisperse bubble carpet rising upwards (cf. Fig.~\ref{fig:pressure_bubbles_snapshots} (d)). Similar behavior was only observed for the second largest electrode with $\Lambda = \SI{60}{\micro\metre}$ (cf. Fig.~\ref{fig:appendix_monodisperse_carpet}). Notably, both electrodes exhibit a bimodal bubble size distribution at lower pressures.

\begin{figure}[ht]
    \centering
    \includegraphics[width=\linewidth]{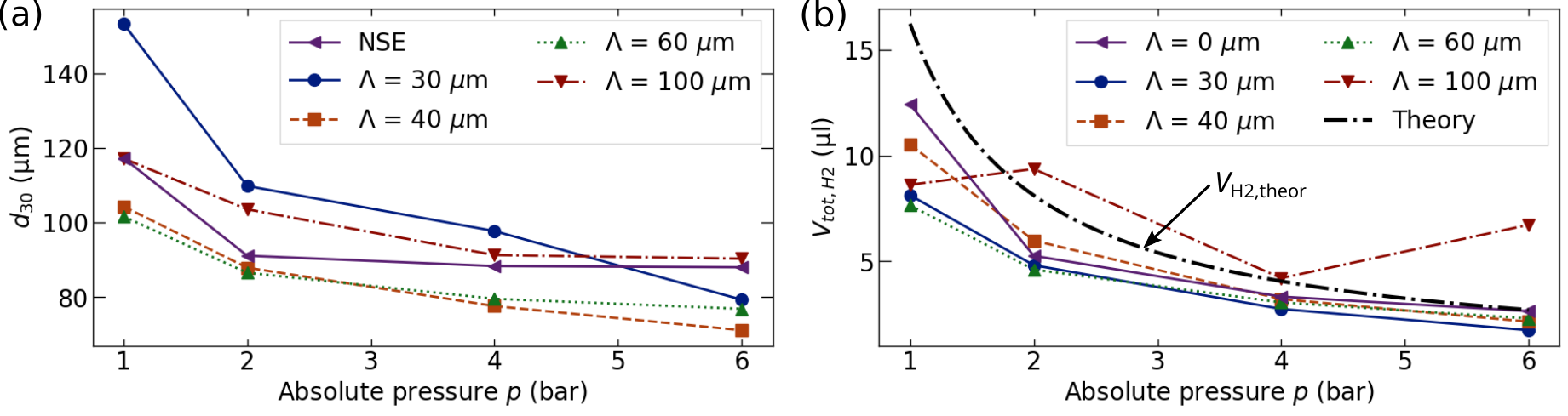}
    \caption{(a) Volumetric mean diameter $d_{30}$ of each size distribution measured as a function of pressure $p$ and electrode surface at \SI{-50}{\milli\ampere\per\centi\metre\squared}. (b) Comparison of measured \ce{H2} volume $V_\mathrm{tot,\ce{H2}}$ of all segmented bubbles with the theoretical \ce{H2} volume $V_\mathrm{\ce{H2}, theor}$ calculated using Eq.~\ref{eq:theor_gas_volume} at \SI{-50}{\milli\ampere\per\centi\metre\squared}. The lines serve only to guide the reader's eye.}
    \label{fig:pressure_bubbles}
\end{figure}

In general, for most measurements the initial pressurization step from 1 to \SI{2}{\bar} shows the largest impact on the bubble size. This can be seen in particular for the current density of \SI{-50}{\milli\ampere\per\centi\metre\squared} (cf. Fig.~\ref{fig:pressure_bubbles} (a)). Similar behavior has been reported by \citeauthor{Liang2024} \cite{Liang2024}.  

As shown in Fig.~\ref{fig:pressure_bubbles_snapshots} (b), the StarDist model cannot segment all the bubbles on these complex structures. However, due to the high frame rate of \SI{1000}{\hertz} each bubble was captured multiple times. To avoid counting bubbles multiple times, their movement was tracked, and the arithmetic mean diameter of all segmentations was calculated. Even if individual bubbles are not detected in individual images, the bubble size distribution still holds. In addition, a two-step validation process was carried out. First, the trained model was validated using a separate validation dataset with regard to the bubble size distribution (cf. Fig.~\ref{fig:appendix_stardist_validation}). Second, the measured total bubble volume was compared with the calculated theoretical \ce{H2} volume. 

According to Henry's law the equilibrium saturation concentration $c_\mathrm{sat}$ increases proportional to the partial pressure. However, as the camera was triggered at steady state ($t \approx \SI{200}{\second}$) it can be assumed that the electrolyte nearby the electrode surface reached already a supersaturation level. Therefore, by combining Faraday's law with the ideal gas law the theoretical produced \ce{H2} volume at a given time $t$ can assumed to be
\begin{align}
    V_\mathrm{\ce{H2}, theor} = \frac{j A_\mathrm{e, theor} t}{z F} \cdot \frac{\mathscr{R}T }{p}. \label{eq:theor_gas_volume}
\end{align}
Here, the ambient temperature $T$ was set to \SI{293.15}{\kelvin} and the pressure $p_0$ to \SI{1}{\bar}. Furthermore, $A_\mathrm{e, theor}$ was assumed to be the cropped image area and additionally all electrode area below it due to the rising bubbles captured by the cameras. This sums up to $\approx$ \SI{21}{\milli\metre\squared}. By plotting this theoretical volume of produced \ce{H2} during the image recording ($t=\SI{12}{\second}$) and comparing it with the actual detected \ce{H2} from the bubble size distribution, the quality of the image analysis can be controlled (see Fig.~\ref{fig:pressure_bubbles} (b)). Therefore, the Laplace pressure $\Delta p = \frac{2\gamma}{r_\mathrm{B}}$, with the bubble interface surface tension $\gamma$ set to \SI{0.072}{\newton\per\metre} \cite{Liang2024}, must be accounted for in the calculation of the \ce{H2} volume from the bubble size distribution assuming perfect spheres $\left(V_\mathrm{B} = \frac{4}{3}\pi r_\mathrm{B}^3\right)$. Furthermore, it is assumed that the bubbles consist out of pure \ce{H2} and thus, water vapor is neglected.
By summarizing all individual bubble volumes, the total volume $V_\mathrm{tot,\ce{H2}}$ of produced \ce{H2} gas can be calculated. As shown in Fig.~\ref{fig:pressure_bubbles} (b), the agreement between $V_\mathrm{\ce{H2}, theor}$ and the total measured bubble volume $V_\mathrm{tot, \ce{H2}}$ is reasonable. This confirms the quality of the applied image analysis.

\subsection*{Influence of pressure on electrode potential}
\begin{figure}[ht]
    \centering
    \includegraphics[width=0.5\linewidth]{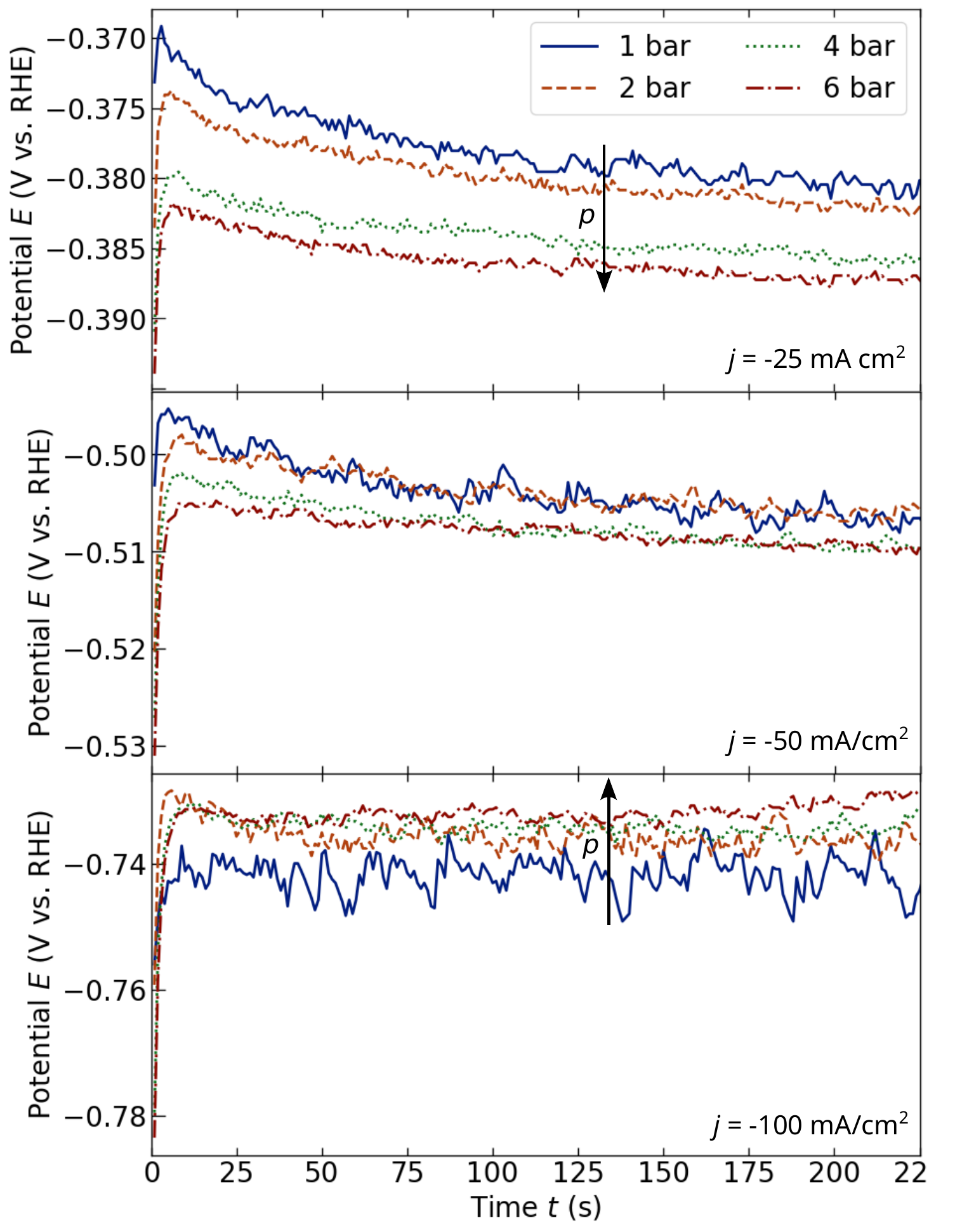}
    \caption{Electrode potential $E$ over time $t$ at all studied current densities $j$ for an example electrode with a spatial period $\Lambda$ of \SI{40}{\micro\metre} featuring a switching point at \SI{-50}{\milli\ampere\per\centi\metre\squared}.}
    \label{fig:pressure_galvanostatic}
\end{figure}

Until now, the effect of pressure on detached bubble sizes has been discussed. It was found that an increase in pressure significantly reduced the bubble size and the width of the bubble size distribution. Since the bubbles significantly influence the electrode potential, the change in bubble size is also reflected in the electrode potential $E$, shown in Fig.~\ref{fig:pressure_galvanostatic}. In general, strong fluctuations in $E$ can be attributed to large bubbles growing and detaching from the electrode surface \cite{Angulo2022}. Thus, the shown fluctuations in the blue curve in Fig.~\ref{fig:pressure_galvanostatic} at $j=\SI{-100}{\milli\ampere\per\centi\metre\squared}$ and $p=\SI{1}{\bar}$ indicate a quite strong influence of the bubbles caused by larger bubbles. Increasing the absolute pressure promotes more homogeneous bubble growth at smaller size. This results in a more uniform distribution of bubble sizes and reduced fluctuations in the electrode potential, as can be seen by comparing the other curves plotted in the same subfigures.

However, as mentioned in the introduction, thermodynamic voltage losses are also affected by absolute pressure, as described by the reformulated Nernst equation in Eq.~\ref{eq:theory_td_losses}. From a theoretical point of view, thermodynamic voltage losses therefore increase with rising pressure. Thus, the electrode potentials at the lowest current density of $j=\SI{-25}{\milli\ampere\per\centi\metre\squared}$ follow the expected theoretical trend of increased potentials with increased pressure (cf. Fig.~\ref{fig:pressure_galvanostatic}). However, when increasing the current density to \SI{-100}{\milli\ampere\per\centi\metre\squared} this behavior switches and the absolute potential is decreasing with increasing pressure. To the best of the author's knowledge, this has never been reported in the literature.



For a better understanding of this phenomenon, the pressure-induced voltage losses $E_\mathrm{PL}$ are now discussed in more detail. For the experimental data these were therefore defined as the difference between the quasi-steady state potential $E_\mathrm{SS}$ at a specific pressure $p$ and $E_\mathrm{SS, ref}$ at the reference case of $p_0  = \SI{1}{\bar}$: 
\begin{align}
    E_\mathrm{PL}(p) = E_\mathrm{SS}(p) - E_\mathrm{SS, ref}(p_0) \label{eq:losses_exp}
\end{align}
Thus, a positive $E_\mathrm{PL}$ indicates that an increase in absolute pressure positively affects the electrode potential relative to $E_\mathrm{theor}$, since $E_\mathrm{theor}$ represents theoretical thermodynamic voltage losses, which always increase the absolute electrode potential. It should be noted that it is not possible to distinguish between thermodynamic and bubble-induced voltage losses in the experimental data. This is because bubble-induced voltage losses also change due to fluctuations in bubble size and electrode coverage.

\begin{figure}[ht]
    \centering
    \includegraphics[width=\linewidth]{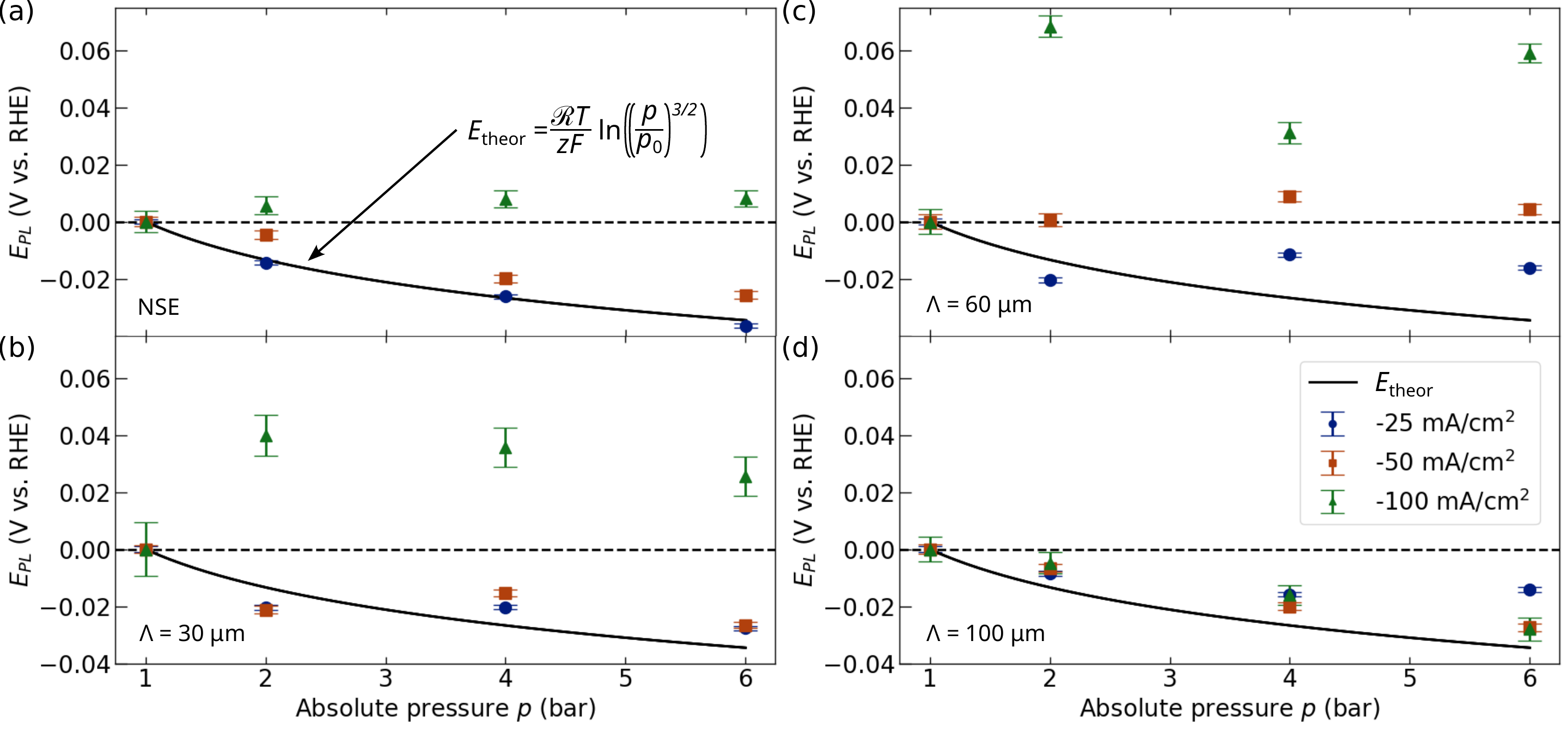}
    \caption{Comparison of the pressure-induced losses $E_\mathrm{PL}$ (see Eq.~\ref{eq:losses_exp}) measured at steady state at different current densities with the theoretical pressure-induced thermodynamic losses $E_\mathrm{theor}$ (see Eq.~\ref{eq:theory_td_losses}) caused by the increase in absolute pressure $p$ for (a) the non-structured electrode (NSE), (b) $\Lambda = \SI{30}{\micro\metre}$, (c) $\Lambda = \SI{60}{\micro\metre}$, and (d) $\Lambda = \SI{100}{\micro\metre}$.}
    \label{fig:pressure_losses}
\end{figure}

At the lowest current density of  $j= \SI{-25}{\milli\ampere\per\centi\metre\squared}$, the calculated $E_\mathrm{PL}$ corresponds fairly well to the theoretical value of $E_\mathrm{theor}$ (cf. Fig.~\ref{fig:pressure_losses}). The agreement is particularly remarkable for the NSE at $j= \SI{-25}{\milli\ampere\per\centi\metre\squared}$. However, when increasing the current density to \SI{-100}{\milli\ampere\per\centi\metre\squared}, most electrodes show a shift towards positive values in $E_\mathrm{PL}$. Thus, an increase in pressure reduces losses and, consequently, lowers the absolute electrode potential. Surprisingly, the largest electrode structure with $\Lambda = \SI{100}{\micro\metre}$ and thus, larger bubbles, shows nearly no influence of the current density. Here, $E_\mathrm{PL}$ follows the theoretical value of $E_\mathrm{theor}$ for all experiments.

According to Faraday's law ($n = I/(zF))$ and the calculation of applied current density using outer geometric area, the total amount of produced \ce{H2} should be the same for all electrodes at a given current density. Even though, the electrodes differ in electrochemically active electrode area, this does not explain the reported phenomenon in Fig.~\ref{fig:pressure_losses}. Consequently, bubble-induced losses play a decisive role here.  The electrode with $\Lambda = \SI{40}{\micro\metre}$ was omitted in Fig.~\ref{fig:pressure_losses} for better readability, as the effect of pressure was not as dominant there (cf. Fig.~\ref{fig:appendix_pressure_losses}).

\subsection*{Dimensional analysis}
Due to the multitude of influencing parameters and reported effects, the Buckingham $\Pi$-theorem was used to derive dimensionless numbers through dimensional analysis. These enable preliminary models of the relationship between bubble-induced and thermodynamic voltage losses as a function of absolute pressure. In total, three dimensionless numbers were obtained and are discussed in the following. Because the NSE does not have a clearly defined structure size parameter, NSE is neglected in the following. Details on the derivation of the dimensionless numbers are provided in the supplementary information (see Sec.~\ref{sec:appendix_dimensional}).

The first dimensionless number $\Pi_\mathrm{1}$ (see Eq.~\ref{eq:dimensionless_pressure}) can be understood as the ratio of the pressure and capillary forces. However, since all parameters have fixed values, the linear relationship shown in Fig.~\ref{fig:appendix_dimensionless_numbers} is non-surprising. Nevertheless, $\Pi_\mathrm{1}$ is used later as dimensionless pressure.
\begin{align}
    \Pi_1 &=\frac{\Lambda p}{\gamma} \label{eq:dimensionless_pressure}
\end{align}

The second dimensionless number $\Pi_\mathrm{2}$ (see Eq.~\ref{eq:dimensionless_bubble_size}) is the relative bubble size and is defined as the ratio of the volumetric mean diameter $d_\mathrm{30}$ over the spatial period $\Lambda$:
\begin{align}
    \Pi_2 &=\frac{d_{30}}{\Lambda} \label{eq:dimensionless_bubble_size}
\end{align}

\begin{figure}[ht]
    \centering
    \includegraphics[width=0.65\linewidth]{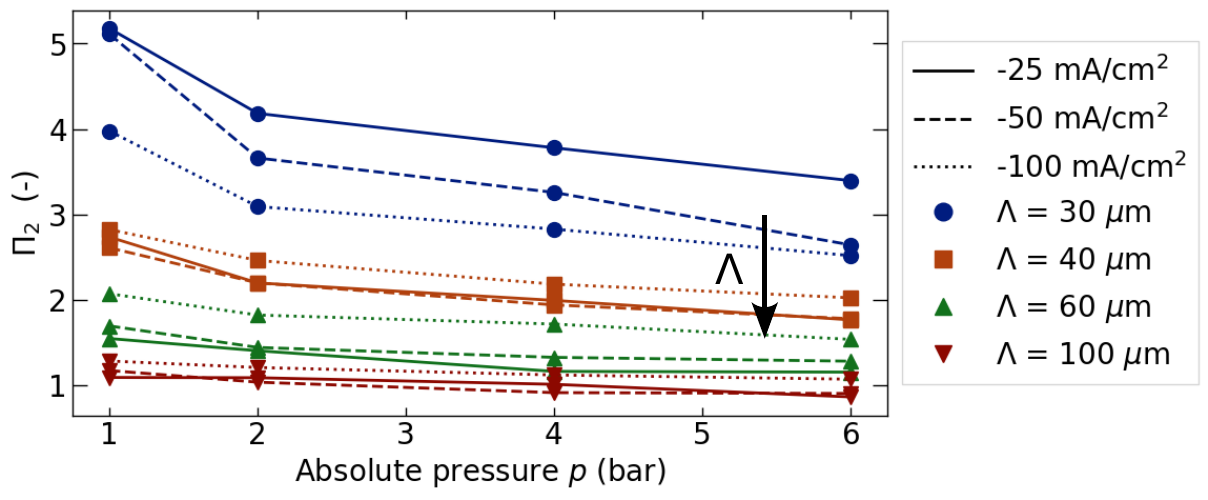}
    \caption{Influence of the absolute pressure on the relative bubble size $\Pi_2$.}
    \label{fig:dimensionless_numbers}
\end{figure}

For $\Lambda = \SI{100}{\micro\metre}$ the ratio of $d_{30}/\Lambda$ remains constant at $\approx 1.3$ to $1.5$ for all studied current densities, while for electrodes with $\Lambda = \SI{40}{\micro\metre}$ and \SI{60}{\micro\metre} $\Pi_\mathrm{2}$ tends to increase with increasing current density (cf. Fig.~\ref{fig:dimensionless_numbers}). Since the spatial period $\Lambda$ remains constant for all electrodes, this means that the volumetric mean diameter $d_{30}$ increases with increasing current density for these two electrodes. The opposite and most prominent behavior is visible for the most hydrophobic electrode with $\Lambda = \SI{30}{\micro\metre}$. Here, the volumetric mean diameter $d_{30}$ decreases with increasing current density.

Even more importantly, for all electrodes except the one with  $\Lambda = \SI{100}{\micro\metre}$, the bubble size decreases with increasing pressure. Thus, bubble-induced losses are reduced. Furthermore, it is again visible that for these electrodes the initial pressurization step from 1 to \SI{2}{\bar} shows the largest impact on the bubble size. Even though it is not directly apparent in this illustration, it should be noted that for the highest current density of \SI{-100}{\milli\ampere\per\centi\metre\squared}, the bubble size is largest for $\Lambda = \SI{100}{\micro\metre}$ and smallest for $\Lambda = \SI{30}{\micro\metre}$ at increased pressures (cf. Fig.~\ref{fig:appendix_bubble_size} (b)).

Based on the current understanding of the bubble growth on these surfaces, the growing bubbles are pinned to the hydrophobic pillars. Thus, for a pillar size of $\Lambda = \SI{100}{\micro\metre}$ and a detached bubble size $d_\mathrm{B}$ ranging from $\approx 100$ to \SI{175}{\micro\metre}, it can be assumed that one bubble is presumably blocking one pillar. Therefore, $\Pi_\mathrm{2}$ may provide an initial estimate of the bubble-induced losses. 

However, since many bubbles captured by the cameras are already detached, calculating or estimating the bubble-induced losses is not possible. As pointed out by \citeauthor{Lake2025}, taking the projected area of the bubbles to estimate the bubble-induced losses leads to non-physical results \cite{Lake2025}. Thus, the actual contact area of the bubble must be known to enable the calculation of bubble-induced losses. However, this is not possible using the available images due to the three-dimensional laser-structured surface and unclear contact line. Therefore, more studies must be performed to reveal the underlying mechanisms of the bubble growth.

\begin{figure}[!t]
    \centering
    \includegraphics[width=0.65\linewidth]{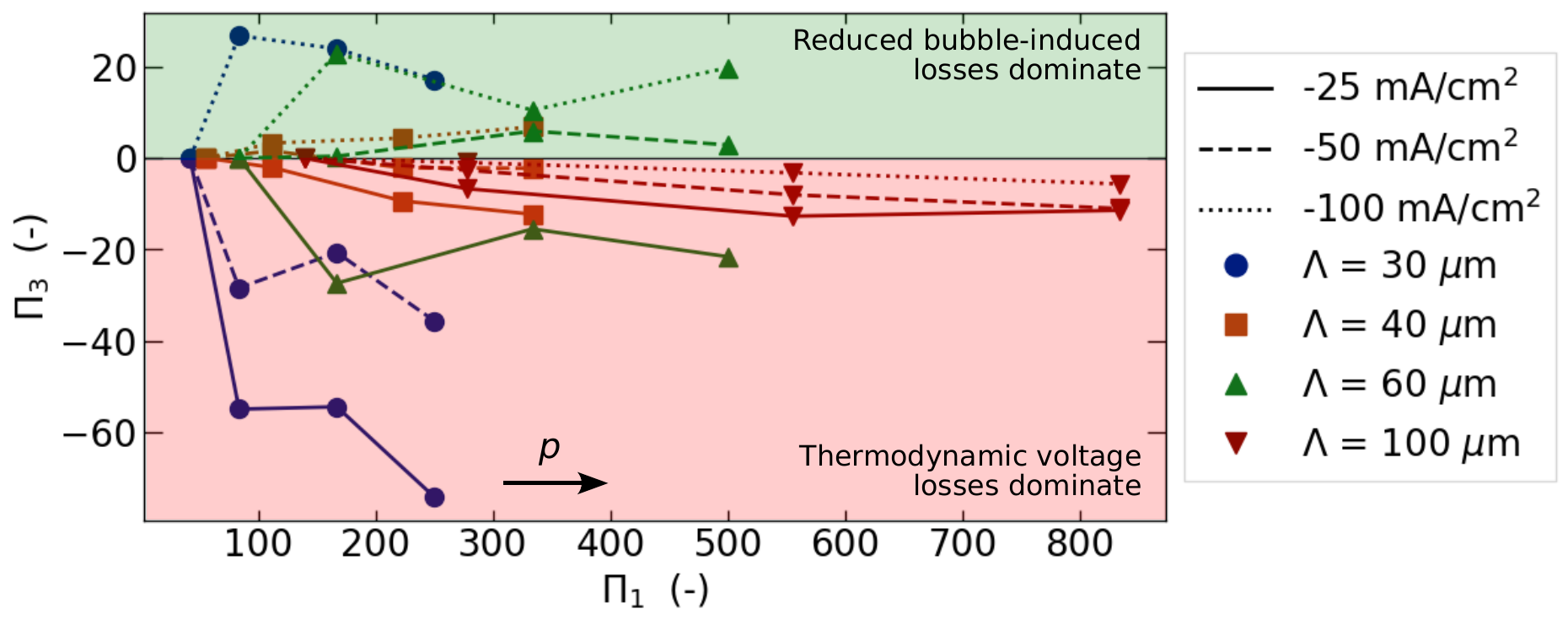}
    \caption{Ratio of resistances of transport of charges through electrolyte and blockage by bubbles, described via $\Pi_3$, over the dimensionless pressure $\Pi_1$.}
    \label{fig:dimensionless_potential}
\end{figure}

The final dimensionless number $\Pi_3$ can be understood as a ratio of two resistances, namely the resistance caused by charge transfer through the electrolyte and the resistance caused by bubbles, since $d_\mathrm{B} = f(\Lambda)$ (cf. Fig.~\ref{fig:dimensionless_numbers}). Consequently, $\Pi_3$ shows similarities to the Wagner number $Wa$. Here, conductivity $\kappa$ and pressure-induced voltage losses $E_\mathrm{PL}$ were set in relation to spatial period $\Lambda$ and current density $j$:
\begin{align}
    \Pi_3 = \frac{-\left(\kappa E_\mathrm{PL}\right)}{\Lambda j} \label{eq:pi_3}
\end{align}
By plotting $\Pi_3$ over $\Pi_1$ the keys results can be summarized in Fig.~\ref{fig:dimensionless_potential}. Multiplying by -1 in the numerator of Eq.~\ref{eq:pi_3} serves only to ensure uniform notation according to Fig.~\ref{fig:pressure_losses}. Thus, a positive value of $\Pi_3$ indicates an improvement of the electrode potential by increased pressure. However, $\Pi_3 > 0$ only occurs in case of 
\begin{enumerate}
    \item high current densities ($|j| \ge \SI{50}{\milli\ampere\per\centi\metre\squared}$) with strong bubble formation and thus, high void fractions, and,
    \item electrodes where the increase in pressure results in a decrease of the bubble sizes.
\end{enumerate}
Therefore, it can be derived that the reported improvement in electrode potential with increasing pressure results from reduced bubble-induced losses. These then compensate for the unavoidable thermodynamic voltage penalty.

\section{Conclusions}
\label{sec:conclusions}
In summary, the present study researched how absolute pressure and bubble size affect electrode performance during HER in alkaline water electrolysis. By tailoring the dimensions of laser-induced structures on Ni electrodes using DLW, different bubble sizes were achieved, directly influencing interfacial transport and electrolysis performance. Increasing the spatial period $\Lambda$ of the generated pillars from \SI{30}{\micro\metre} to \SI{100}{\micro\metre} led to larger detached bubbles, especially at increased current densities of \SI{-100}{\milli\ampere\per\centi\metre\squared}. On the contrary, increasing the absolute pressure from 1 to \SI{6}{\bar} reduced the size of the bubbles. Here, the most significant reduction in bubble size appears for most electrodes at the initial pressurization from \SI{1}{\bar} to \SI{2}{\bar}. This also applies for the cathodic potential. However, the direction of the potential shift strongly depends on the applied current density and the balance between bubble-induced and thermodynamic voltage losses. At a low current density of \SI{25}{\milli\ampere\per\centi\metre\squared}, the cathodic potential increases according to the pressure, following the thermodynamic voltage losses given by the Nernst equation. This leads to a maximum increase in potential of $\approx$ \SI{23}{\milli\volt} at a pressure of $p=\SI{6}{\bar}$. Thus, the overall electrolysis efficiency is hindered by elevated pressures. When increasing the current density a switch is observed resulting in a decrease in the cathodic potential by up to $\approx$ \SI{60}{\milli\volt} with increasing pressure. This has been proven for all electrodes except the one with the largest spatial period of pillar patterns and, thus, the largest detached bubbles. Consequently, elevated pressures at high current operation promote the reduction of bubble-induced losses, which then outweigh the unavoidable thermodynamics voltage losses. Therefore, the pressurization enhances the overall electrolysis performance.

\section{Material and methods}    
\label{sec:methods}
\subsection*{Electrode fabrication}
For the laser-structuring, \SI{200}{\micro\metre} thick pure nickel (\SI{99.95}{\%} purity, Goodfellow, USA) sheets with the dimensions of $30 \times \SI{30}{\milli\metre\squared}$ were used. The samples were delivered in rolled condition with an average surface roughness $S_\text{a}$ of \SI{0.19}{\micro\metre}. Prior to the laser-texturing, the samples were cleaned with ethanol. Following the laser process, the samples were stored under atmospheric conditions and without any other additional treatment.

The laser-texturing process was carried out using a picosecond slab-shaped solid-state laser (Edgewave InnoSlab, Würselen, Germany) emitting at a wavelength $\lambda$ of \SI{532}{\nano\metre} and with a pulse width of \SI{12}{\pico\second}. The laser beam is guided by high reflective mirrors towards a beam expander and further to a 2D-galvo scanner (Raylase Super Scan III, Weßling, Germany). Furthermore, a f-theta lens is applied with a focal length of \SI{100}{\milli\metre}, leading to a beam diameter in the focal plane of \SI{20}{\micro\metre}. The setup for laser-texturing is shown in Fig.~\ref{fig:appendix_ls_setup} in the -supplementary information.

\subsection*{Experimental methods}

\subsubsection*{Characterization of electrode surface}
For evaluating the surface topography of the laser-structured samples, White Light Interferometric images (Sensofar S-Neox, Spain) were recorded by employing 50x magnification objective. The surface profiles and average structure depth values were obtained using the SensoMAP Advanced Analysis Software (Sensofar, Spain). The static water contact angle was measured by applying a water droplet with a volume of \SI{2}{\micro\liter} using a contact angle measurement system (OCA 200, DataPhysics Instruments GmbH, Germany). All working electrodes (WE) were cleaned in a ultrasonic bath with Isopropanol for \SI{5}{\minute}, rinsed with deionized (DI) water and stored in DI water. Before each measurement performed in the electrochemical cell, the WE was cleaned with ethanol and subsequently rinsed with DI water to remove any remaining contamination before mounting them onto the electrode holder.

Monochromatic Al K$\alpha$ (\SI{1486.6}{eV}) X-rays were focused to a \SI{100}{\micro\metre} spot using a PHI VersaProbeII Scanning XPS system (ULVAC-PHI). The photoelectron takeoff angle was \SI{45}{\degree} and the pass energy in the analyzer was set to \SI{117.50}{eV} (\SI{0.5}{eV} step) for survey scans and \SI{46.95}{eV} (\SI{0.1}{eV} step) to obtain high energy resolution spectra  for the C 1s, O 1s, N 1s, Si 2p, S 2p, Ag 3d, Fe 2p, Cr 2p, Ca 2p, K 2p and Ni 2p regions. A dual beam charge compensation with \SI{7}{eV} Ar+ ions and \SI{1}{eV} electrons was used to maintain a constant sample surface potential regardless of the sample conductivity. All XPS spectra were charge referenced to the unfunctionalized, saturated carbon (C-C) C 1s peak at \SI{285.0}{eV}. The operating pressure in the analytical chamber was less than \SI{3e-9}{\milli\bar}. Deconvolution of spectra was carried out using PHI MultiPak software (v.9.9.3). Spectrum background was subtracted using the Shirley method. More details on the deconvolution and high resolution spectra can be found in Sec.~\ref{sec:appendix_xps} in the supplementary information.

\subsubsection*{Characterization of electrode performance}
The electrodes were mounted vertically in the electrolyzer (see Fig.~\ref{fig:appendix_pressure_setup} in the supplementary information) facing the counter electrode aligned in parallel, which was a \ce{Pt} mesh with a much larger electrode area. Thereby, the cut edges and the electrical connection of the WE were covered using KOH-resistant silicone (101RF, Microset, UK). The electrode area was thus limited to \SI{2}{\centi\metre} $\times$ \SI{2}{\centi\metre}. In contrast to all previous studies \cite{Rox2025b, Rox2025a}, an Ag/AgCl electrode is used as reference electrode here. For better comparability, all measured potentials were converted to \si{\volt} vs. RHE. Before the galvanostatic measurements, a constant OCP was measured and the WE was activated using three cyclic voltammetries from \SI{-0.2}{\volt} to \SI{0.8}{\volt} vs. RHE at a scan rate of $\nu = \SI{0.01}{\volt\per\second}$. In total, three different current densities ($j = -25, -50, \SI{-100}{\milli\ampere\per\centi\metre\squared}$) and four absolute pressures ($p = 1, 2, 4, \SI{6}{\bar}$) were studied at a constant electrolyte concentration of \SI{1}{M} KOH without external flow.

A VersaStat (AMETEK, USA) was used as electrochemical workstation. The absolute pressure $p$ inside the electrolyzer was adjusted using pressurized \ce{N2} and controlling its mass flow rate at the cell outlet via a pressure controller (uncertainty: \SI{0.2}{\%}, Bronkhorst High-Tech, Netherlands). High-speed images of the bubbles were taken using a IDT OS-7 S3 (696 × 672 px, IDT, USA) at a sample rate of \SI{1000}{\hertz} and a spatial resolution of \SI{194}{\px\per\milli\metre}. The camera was triggered as soon as the measured electrode potential reached steady state (at $t \approx \SI{200}{\second}$). For the illumination of the electrode surface an LED-panel (CCS TH2, Japan) was used. Since the electrode surfaces refracted the light in different ways, the exposure time was adjusted between \SI{598}{\micro\second} and \SI{989}{\micro\second} to achieve the best image quality for each WE. 

\subsubsection*{Characterization of bubble evolution}
A machine-learning based approach was chosen to segment the bubbles. Therefore, a StarDist model (v.0.8.5) \cite{Schmidt2018, Weigert2020} was trained with a randomly chosen set of 100 manual labeled images. After segmentation of the bubbles in the images, all objects were linked using trackpy \cite{Allan2023} and finally, blurred bubbles were eliminated by calculating the size-normalized variance of the image Laplacian (Var($\Delta$)$\cdot d_\text{B}$). Therefore, all segmented bubbles below the 1 \% quantile of this metric were excluded from further analysis, as they correspond to blurred bubbles \cite{Rox2023}.

\subsection*{Dimensional analysis}
Neglecting temperature, seven relevant parameters remain, which are shown in Table~\ref{tbl:relevant_parameters}. Since these refer to four fundamental units (M, L, T, I), three dimensionless numbers were obtained using the Buckingham $\Pi$-theorem. Details about the matrix transformations can be found in the supplementary information (see Sec.~\ref{sec:appendix_dimensional}).
\begin{table}[ht]
    \centering
    \caption{Relevant parameters and their SI units.}
    \label{tbl:relevant_parameters}
    \begin{tabular*}{0.7\textwidth}{@{\extracolsep{\fill}}llll}
    \hline
    & \multicolumn{2}{c}{Parameter} & Unit \\
    \hline
    \multirow{4}{*}{Process parameters}  & Absolute pressure & $p$ & \si{\kilo\gram\per{\metre\second\squared}}\\
    & Current density & $j$ & \si{\ampere\per\metre\squared} \\
    & Electrode potential & $E_\mathrm{PL}$ & \si{\kilogram\metre\squared\per{\ampere\cubic\second}} \\
    & Bubble size & $d_{30}$ & \si{\metre} \\
    Electrode & Spatial period & $\Lambda$ & \si{\metre} \\
    \multirow{2}{*}{Material}  & Conductivity & $\kappa$ & \si{\ampere\squared\cubic\second\per{\kilogram\cubic\metre}}\\
    & Surface tension & $\gamma$ & \si{\kilogram\per\second\squared}\\
    \hline
  \end{tabular*}
\end{table}

\section*{Author Contributions}
H.R.: Conceptualization, Investigation, Data curation, Formal Analysis, Methodology, Visualization, Writing - original draft, Editing; F.L.: Investigation, Data curation, Formal Analysis, Methodology,  Writing - review \& editing; R.B.: Investigation,  Data curation, Methodology, Visualization, Writing - review \& editing; M.M.M.: Investigation, Formal Analysis, Writing - review \& editing; K.S.: Investigation, Formal Analysis; X.Y.: Conceptualization, Supervision, Writing - review \& editing; A.F.L.: Supervision, Writing - review \& editing; R.v.d.K.:  Funding acquisition, Project administration, Supervision, Writing - review \& editing; K.E.: Funding acquisition, Project administration, Supervision, Writing - review \& editing.

\section*{Conflicts of interest}
There are no conflicts to declare.

\section*{Data availability}
Data for this article, including electrochemical measurement data, raw images and relevant metadata of the performed experiments are available at RODARE at \href{https://doi.org/10.14278/rodare.4547}{10.14278/rodare.4547}.

\section*{Acknowledgments}
This project is supported by the Federal State of Saxony in terms of the "European Regional Development Fund" (H2-EPF-HZDR), the Helmholtz Association Innovation pool project "Solar Hydrogen", the Hydrogen Lab of the School of Engineering of TU Dresden, and BMBF (project ALKALIMIT, grant no. 03SF0731A). The authors would also like to thank Mengyuan Huang and Gerd Mutschke from the Institute of Fluid Dynamics at the Helmholtz-Zentrum Dresden-Rossendorf for their helpful discussions on the influence of pressure on the force balance of bubbles.


\balance


\bibliography{references} 
\newpage
\setcounter{section}{0}
\setcounter{equation}{0}
\setcounter{figure}{0}
\setcounter{table}{0}
\renewcommand{\thefigure}{S\arabic{figure}}
\renewcommand{\thetable}{S\arabic{table}}
\renewcommand{\thesection}{S\arabic{section}}

\section*{Electronic supplementary information (ESI):\\ Bubble-induced versus thermodynamic voltage losses during pressurized alkaline water electrolysis}
\newpage
\section{Relevant forces on \ce{H2} bubble}
\label{sec:appendix_theory}
\begin{figure}[ht]
    \centering
    \includegraphics[width=0.4\linewidth]{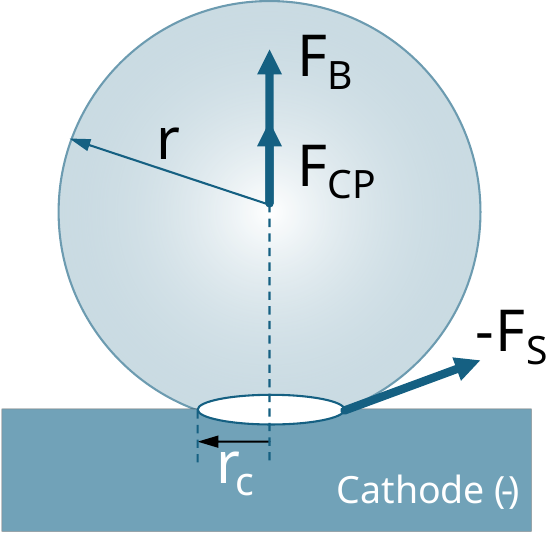}
    \caption{Schematic drawing of the two forces discussed here that act on a growing \ce{H2} bubble, with sketched circular three-phase contact line. Here, $F_\mathrm{S}$ denotes the interfacial tension force and $F_\mathrm{CP}$ denotes the contact pressure force. The buoyancy force $F_\mathrm{B}$ is used to indicate the direction of gravitational acceleration.}
    \label{fig:appendix_bubble_forces}
\end{figure}
\section{XPS Spectra}
\label{sec:appendix_xps}
\subsection{Deconvolution and fitting of XPS Spectra}
High energy resolution spectra were obtained for the C 1s, O 1s, N 1s, Si 2p, S 2p, Ag 3d, Fe 2p, Cr 2p, Ca 2p, K 2p and Ni 2p regions. Deconvolution of spectra was carried out using PHI MultiPak software (v.9.9.3). Spectrum background was subtracted using the Shirley method. In the following, only the spectra mentioned in this work are discussed in more detail.

The C 1s spectra for all samples were fitted with three components. First line centered at \SI{285.0}{eV} arise from aliphatic carbon \ce{C-C}, second line lies at \SI{286.6}{eV} and indicates presence of \ce{C-O} and/or \ce{C-O-C} and/or \ce{C-N} bonds and third line centered at \SI{288.1}{eV} indicates presence of \ce{C=O} and/or \ce{N-C=O} and/or \ce{O-C-O} bonds \cite{beamson1993,rouxhet2011}.

The O 1s spectra were fitted using three lines: first line centered at 529.8 eV which indicates presence of metal oxides (\ce{O-Ni}, \ce{O-Fe}, \ce{O-Cr}), second line at \SI{531.6}{eV} indicates presence of defective oxygen in metal oxides and/or \ce{O=C} and/or \ce{S-O} and/or \ce{O-Si} bonds and the last line found at \SI{533.1}{eV} which can originate either from \ce{O-H} and/or \ce{C-O} type bonds and/or adsorbed \ce{H2O} \cite{beamson1993,genet2008, wagner1982}.

The spectra collected at Ni 2p$_{3/2}$ region are similar for all samples where nickel was detected. Each spectrum was fitted with up to six lines. First line centered at \SI{852.7}{eV} indicates presence of metallic nickel whereas second line found at \SI{854.5}{eV} indicates the \ce{Ni2+} in nickel oxide (\ce{NiO}) and/or hydroxide \cite{biesinger2009,biesinger2011,biesinger2012}. The four lines within energy range of 855 – \SI{866}{eV} are due to the multiplet splitting phenomena.

Spectra collected at Fe 2p$_{3/2}$ region were fitted with up to six components with first line centered at \SI{709.6}{eV} which points out the existence of Fe$^{3+}$ oxidation state. The four lines within energy range of 711 – \SI{714}{eV} are due to the multiplet splitting phenomena and the position of last shake-up line found at $\approx$ \SI{719}{eV} is additional parameter which ensure Fe$^{3+}$ oxidation state of iron in the samples \cite{biesinger2011,Grosvenor2004}.
\newpage
\subsection{High resolution spectra}
\begin{figure}[ht]
    \centering
    \includegraphics[width=0.85\linewidth]{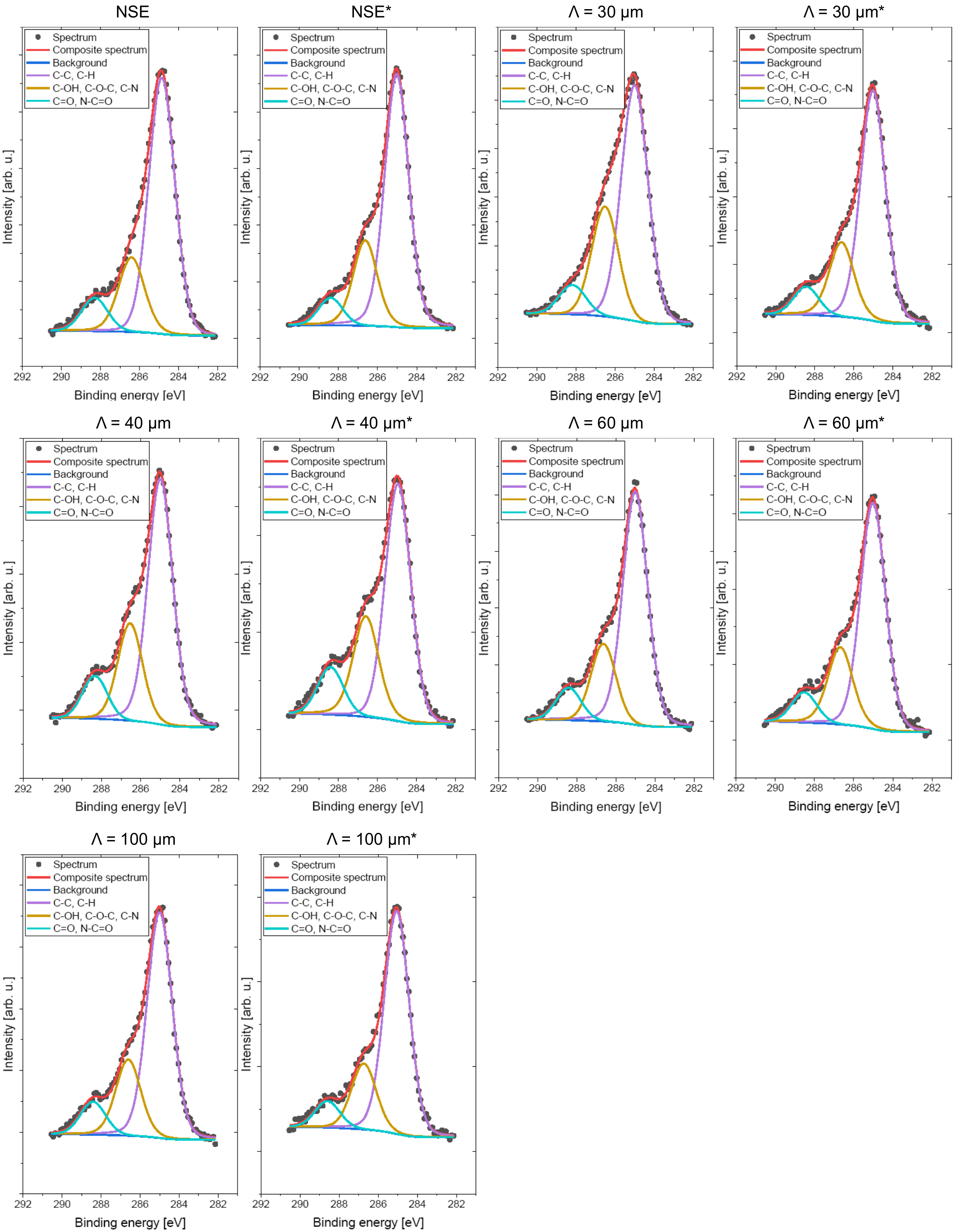}
    \caption{Fitted C 1s spectra of all studied electrodes measured by XPS in comparison to the non-structured electrode (NSE). Annotations: $^\star$Fresh electrode samples without heat treatment.}
    \label{fig:xps_spectra_c}
\end{figure}
\begin{figure}[ht]
    \centering
    \includegraphics[width=0.85\linewidth]{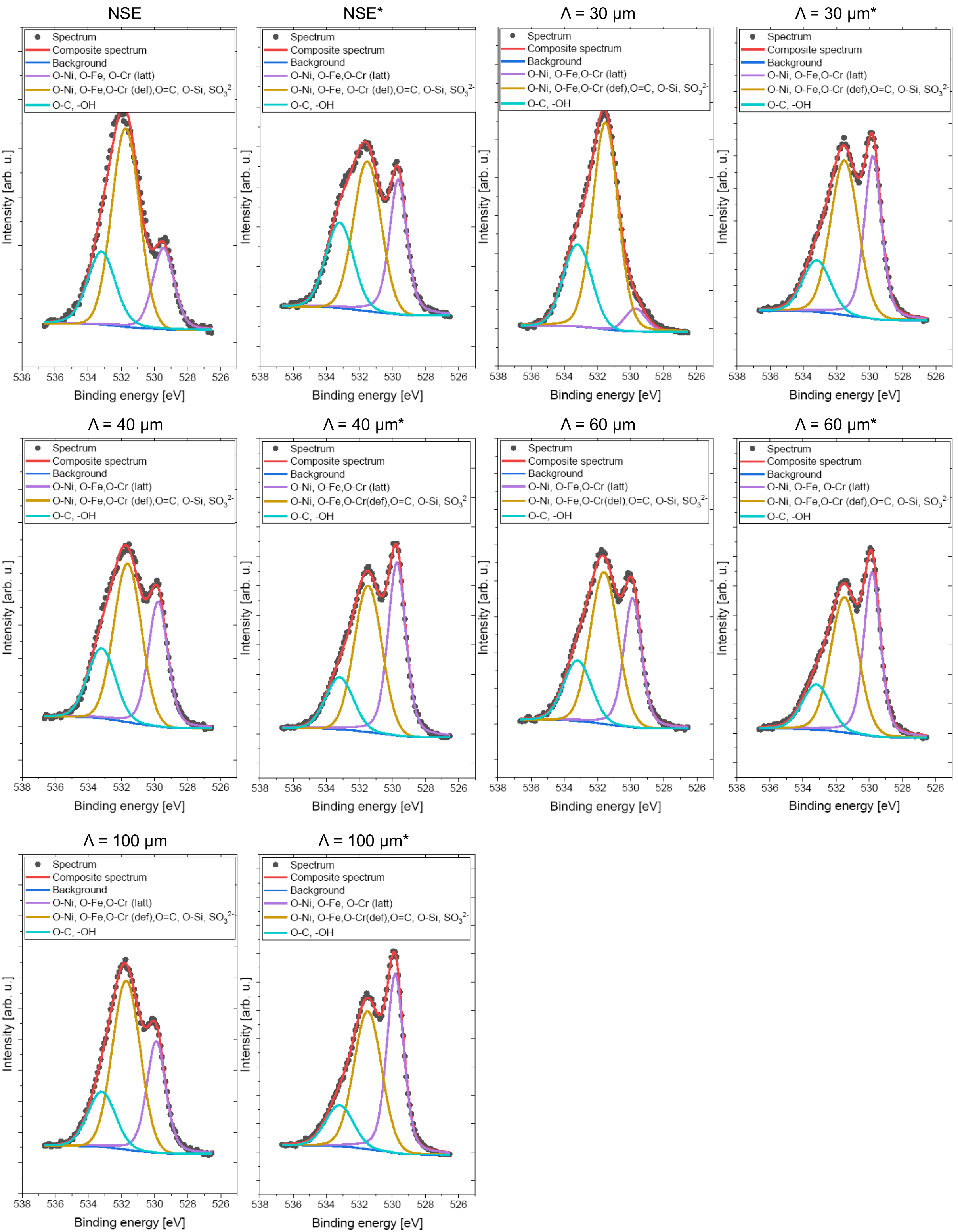}
    \caption{Fitted O 1s spectra of all studied electrodes measured by XPS in comparison to the non-structured electrode (NSE). Annotations: $^\star$Fresh electrode samples without heat treatment.}
    \label{fig:xps_spectra_o}
\end{figure}
\begin{figure}[ht]
    \centering
    \includegraphics[width=0.85\linewidth]{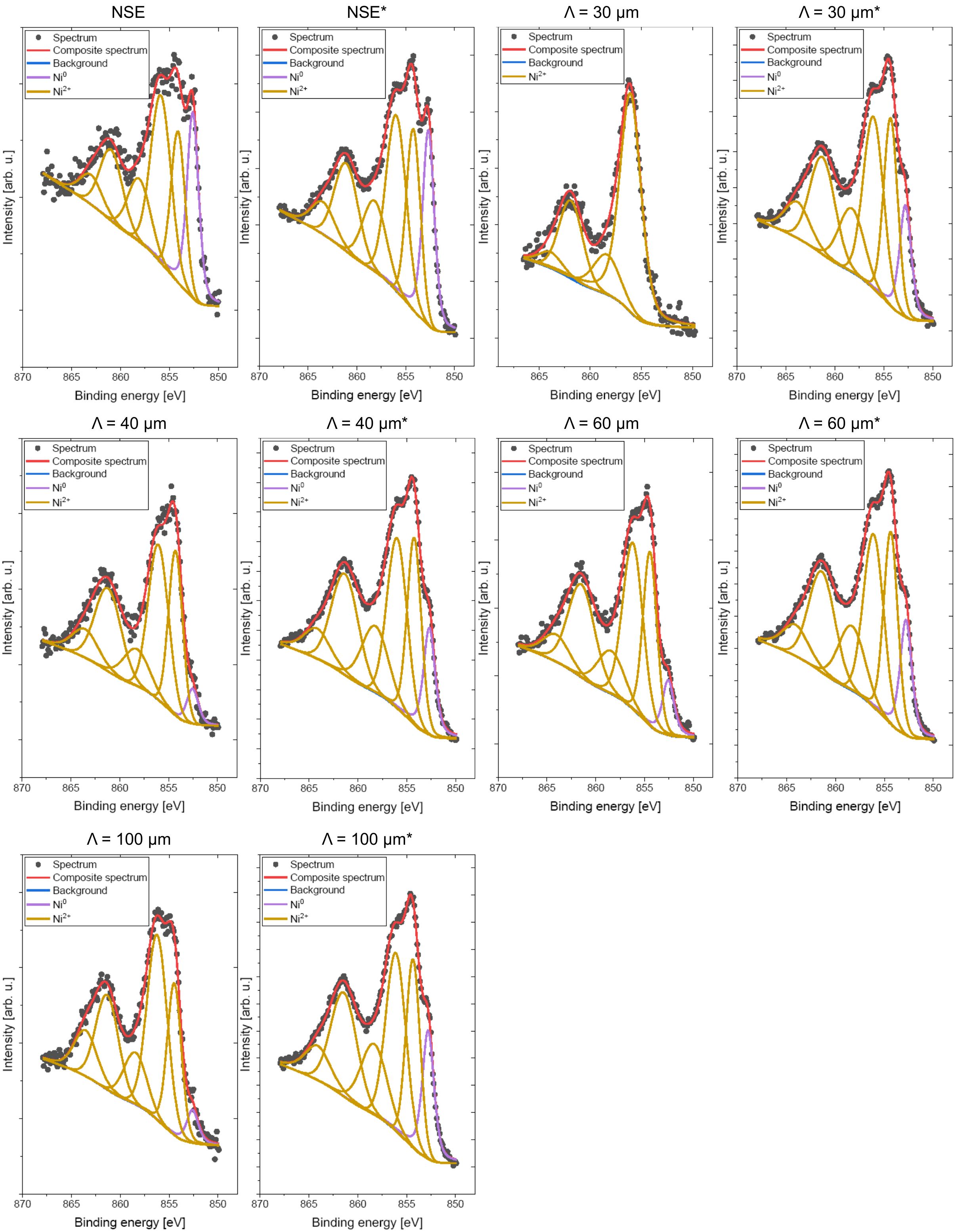}
    \caption{Fitted Ni 2p$_{3/2}$ spectra of all studied electrodes measured by XPS in comparison to the non-structured electrode (NSE). Annotations: $^\star$Fresh electrode samples without heat treatment.}
    \label{fig:xps_spectra_ni}
\end{figure}
\FloatBarrier
\begin{figure}[ht]
    \centering
    \includegraphics[width=0.6\linewidth]{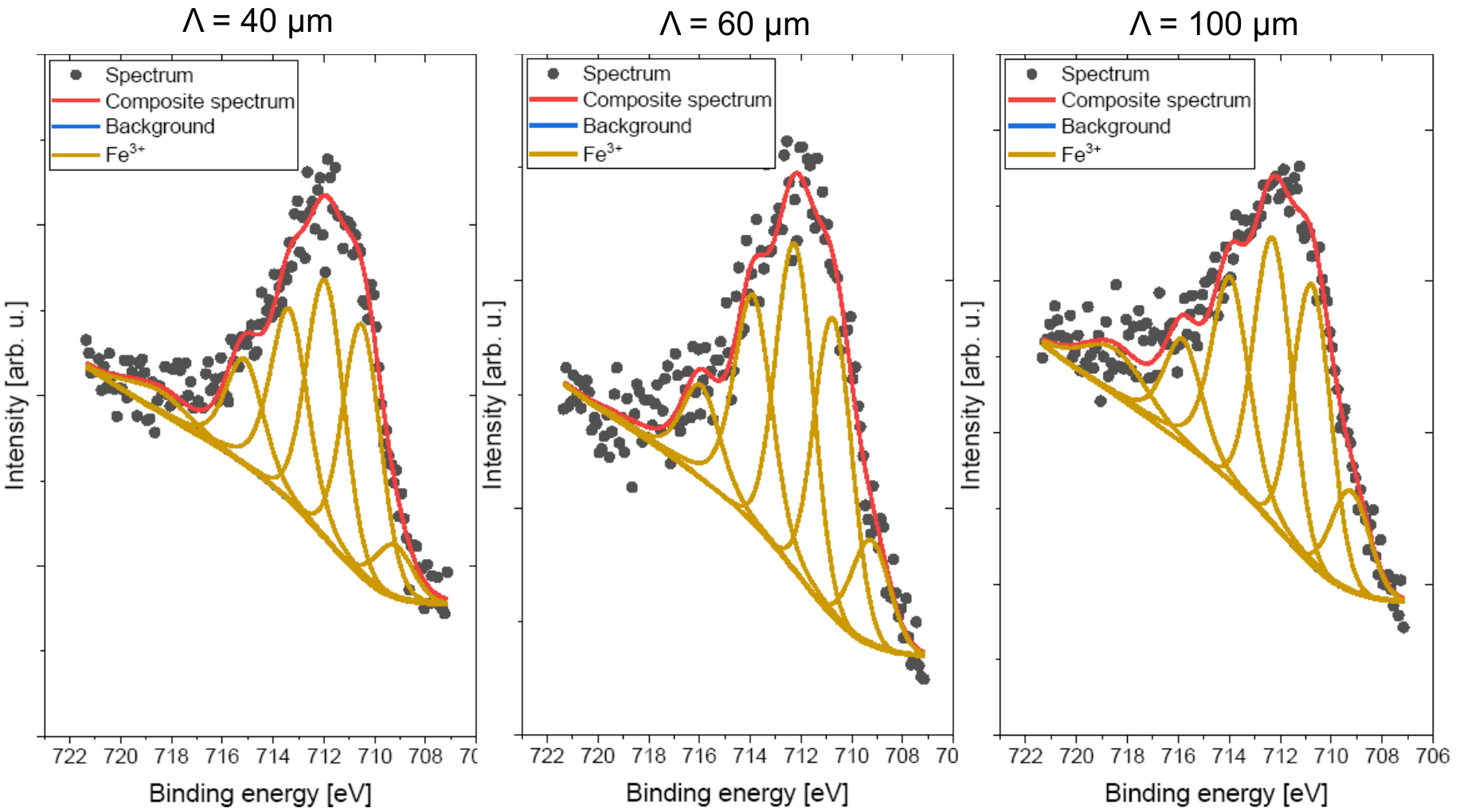}
    \caption{Fitted Fe 2p$_{3/2}$ spectra of all studied electrodes measured by XPS. \textcolor{black}{For a better readability, only electrodes with a measurable Fe$^{3+}$ content are shown (cf. Table~2).}}
    \label{fig:xps_spectra_fe}
\end{figure}

\section{Supporting figures for experimental setups}
\label{sec:appendix_setup}
\begin{figure}[ht]
    \centering
    \includegraphics[width=\linewidth]{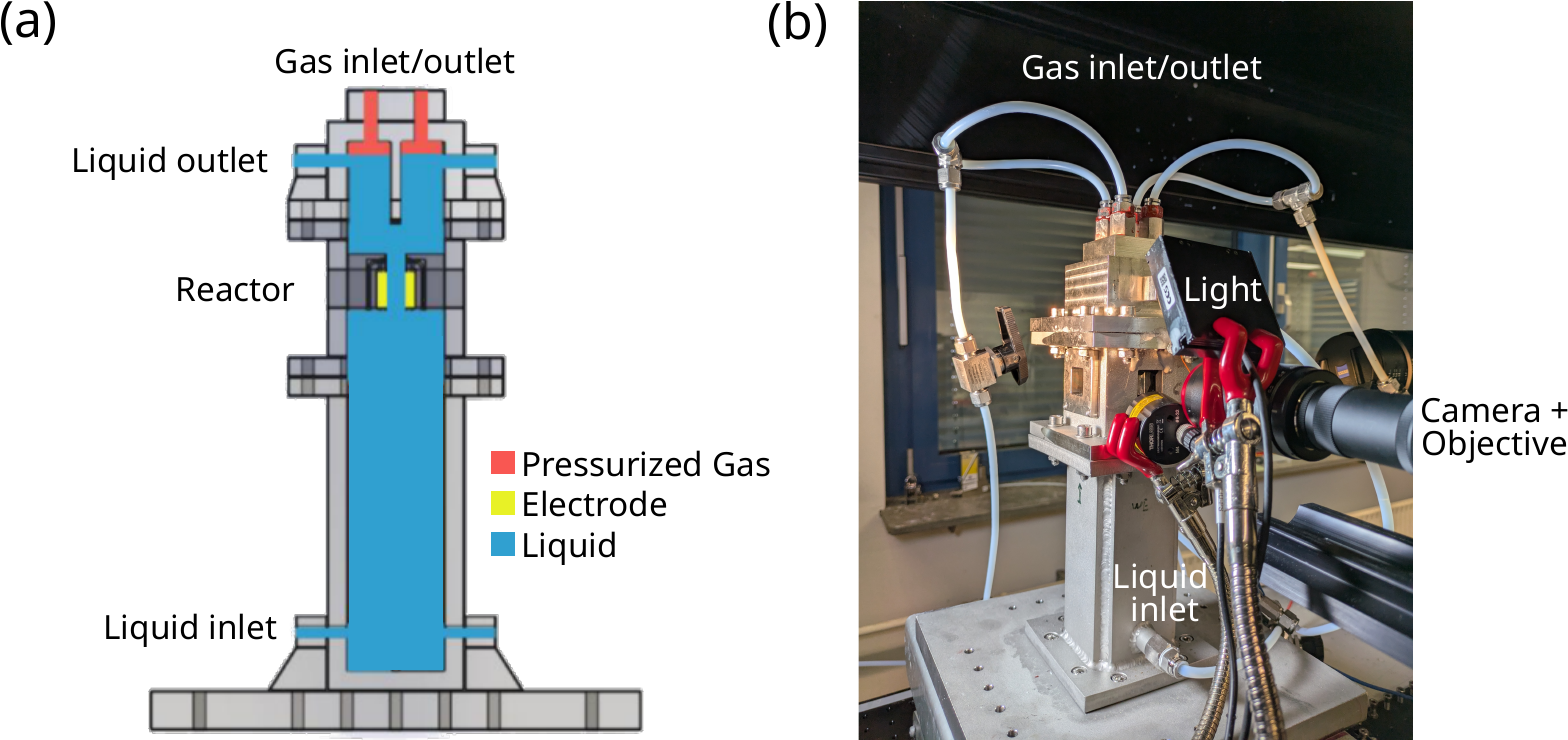}
    \caption{(a) Schematic drawing of the experimental setup at elevated pressures. Adapted from \citeauthor{Liang2024} \cite{Liang2024}. (b) Photography of the pressurized cell with imaging equipment.}
    \label{fig:appendix_pressure_setup}
\end{figure}
\begin{figure}[ht]
    \centering
    \includegraphics[width=\linewidth]{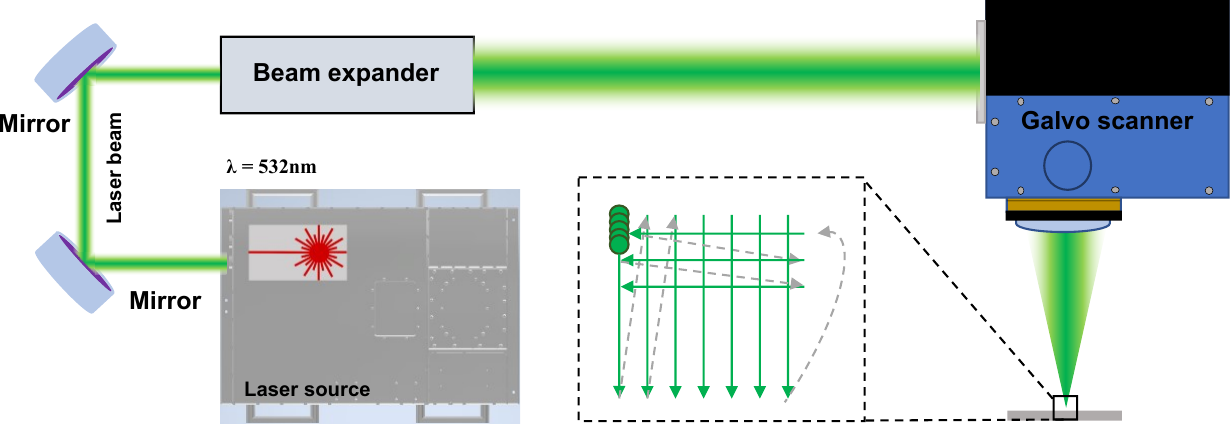}
    \caption{Schematic drawing of laser texturing setup with unidirectional scan strategy to generate a cross-like pattern.}
    \label{fig:appendix_ls_setup}
\end{figure}
\newpage
\section{Validation of image analysis}
\label{sec:appendix_image_analysis}
\begin{figure}[ht]
    \centering
    \includegraphics[width=0.65\linewidth]{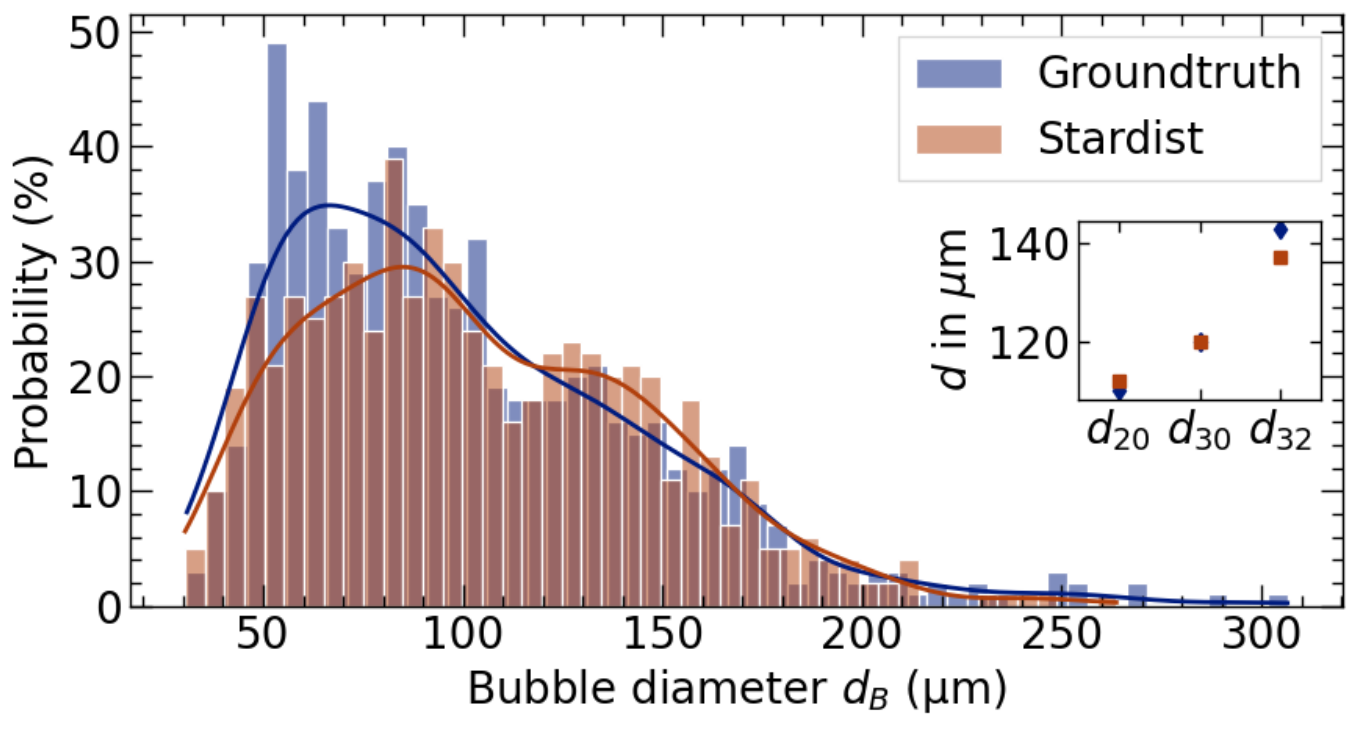}
    \caption{Comparison of bubble size distribution and calculated mean diameters of the used StarDist model in comparison to a validation data set out of 30 randomly chosen images.}
    \label{fig:appendix_stardist_validation}
\end{figure}
\newpage
\section{Supporting figures on influence of pressure}
\begin{figure}[ht]
    \centering
    \includegraphics[width=0.66\linewidth]{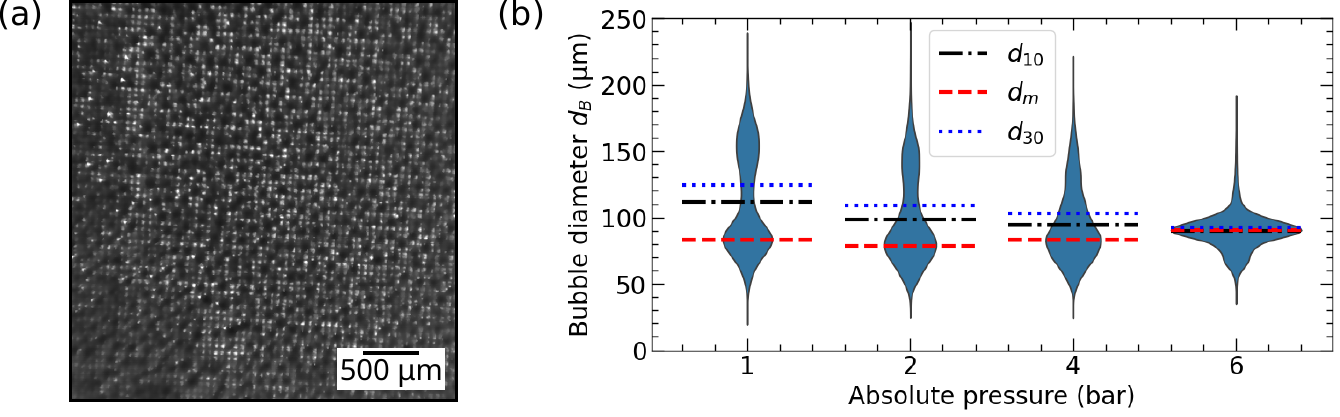}
    \caption{(a) Snapshot of the nearly monodisperse bubble carpet evolving at the electrode bubbles with $\Lambda = \SI{60}{\micro\metre}$ at \SI{6}{\bar} and \SI{-100}{\milli\ampere\per\centi\metre\squared}. (b) Bubble size distributions at \SI{-100}{\milli\ampere\per\centi\metre\squared} for $\Lambda = \SI{60}{\micro\metre}$ showing the shift from bimodal bubble size distributions towards a \textcolor{black}{nearly} monodisperse bubble carpet.}
    \label{fig:appendix_monodisperse_carpet}
\end{figure}
\begin{figure}[ht]
    \centering
    \includegraphics[width=0.87\linewidth]{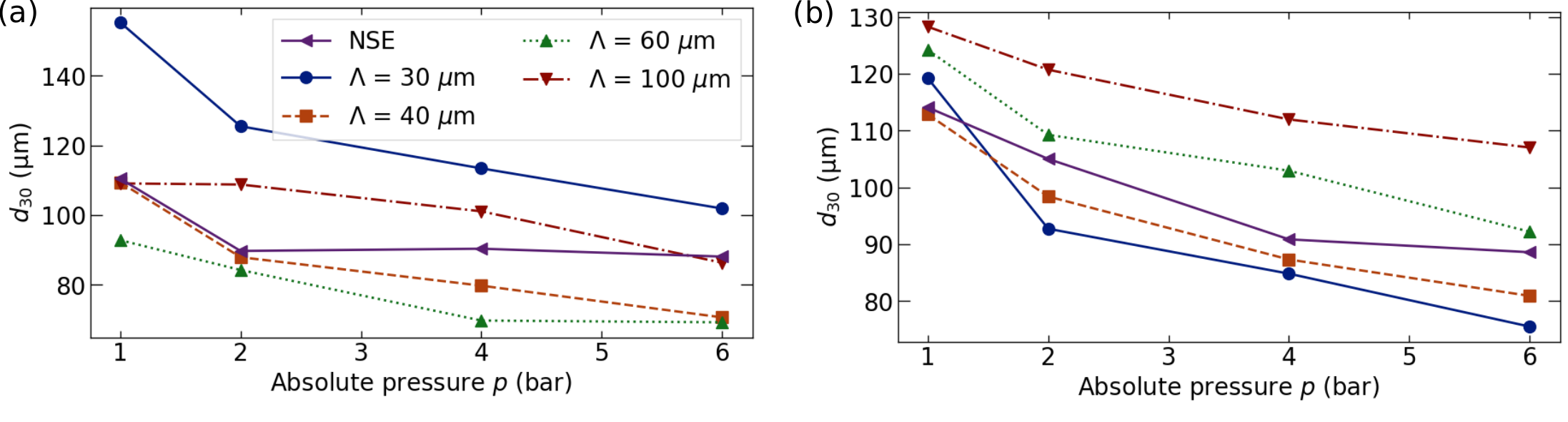}
    \caption{Volumetric mean diameter $d_{30}$ of each size distribution measured as a function of pressure and electrode surface at (a) \SI{-25}{\milli\ampere\per\centi\metre\squared} and (b) \SI{-100}{\milli\ampere\per\centi\metre\squared}.}
    \label{fig:appendix_bubble_size}
\end{figure}
\begin{figure}[ht]
    \centering
    \includegraphics[width=0.63\linewidth]{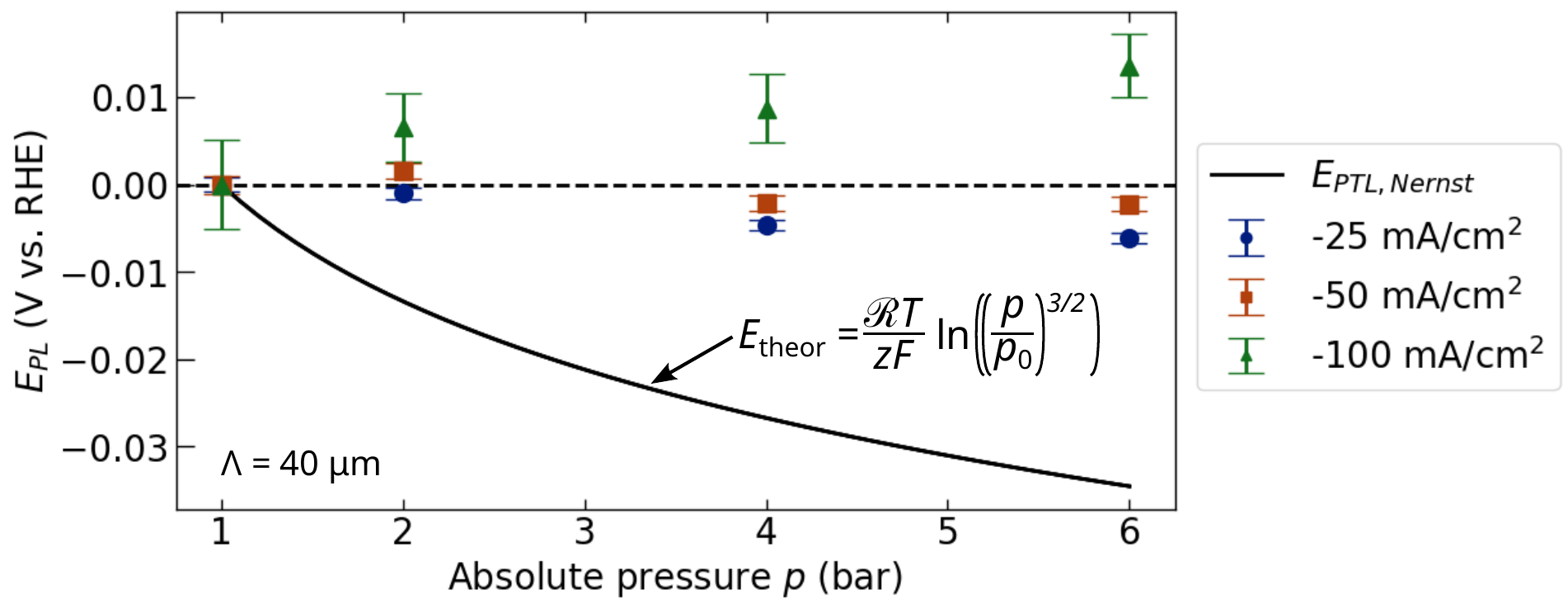}
    \caption{\textcolor{black}{Comparison of the pressure-induced losses $E_\mathrm{PL}$ (see Eq.~8) of the electrode with a spatial period of $\Lambda = \SI{40}{\micro\metre}$ measured at steady state at different current densities with the theoretical pressure-induced thermodynamic losses $E_\mathrm{theor}$ (see Eq.~3) caused by the increase in absolute pressure $p$.}}
    \label{fig:appendix_pressure_losses}
\end{figure}
\newpage
\section{Dimensional analysis}
\label{sec:appendix_dimensional}
\begin{figure}[ht]
    \centering
    \includegraphics[width=0.85\linewidth]{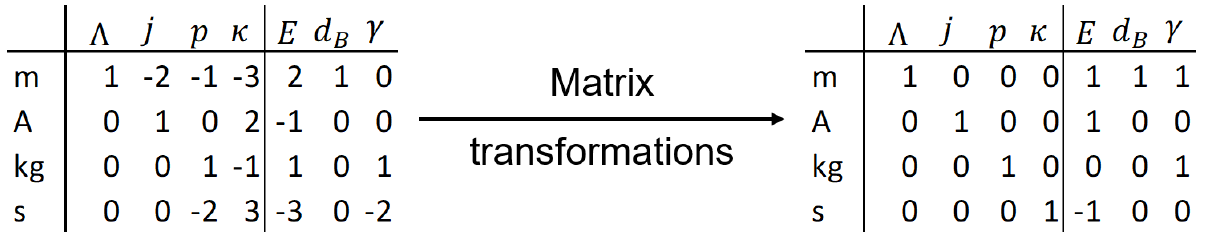}
    \caption{Initial dimension matrix and resulting unit matrix of the kernel matrix with the associated residual matrix.}
    \label{fig:appendix_dimensional}
\end{figure}
\begin{figure}[ht]
    \centering
    \includegraphics[width=0.6\linewidth]{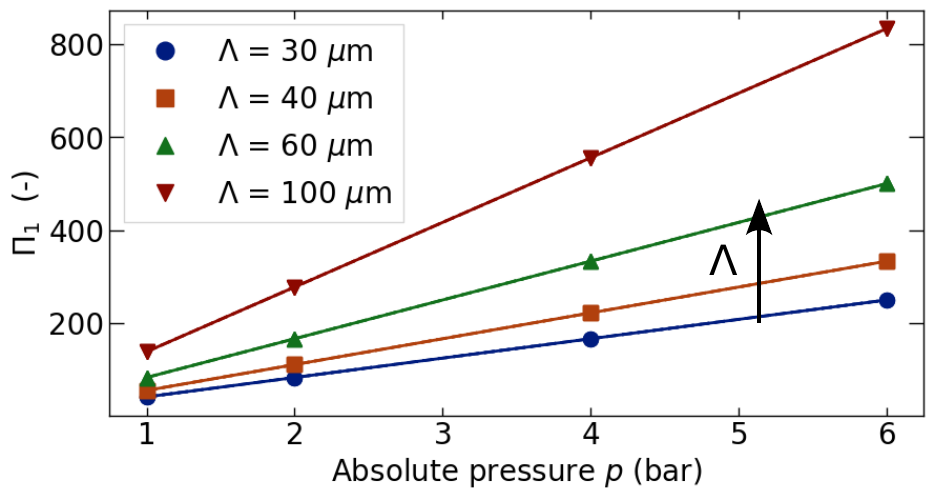}
    \caption{Influence of the absolute pressure on the dimensionless pressure $\Pi_1$ (ratio of pressure and capillary forces).}
    \label{fig:appendix_dimensionless_numbers}
\end{figure}
\FloatBarrier
\section{Data and videos}
Sample data sets with raw images, electrochemical measurement data, and results can be found at \href{https://doi.org/10.14278/rodare.4547}{10.14278/rodare.4547}. Due to the size of the complete data, the remaining image data can be made available upon request.

The provided characteristic videos are named after following scheme:
\begin{center}
  \textit{SpatialPeriod\_Pressure\_CurrentDensity} \\
    \textit{E.g.: 100$\mu$m\_6bar\_-100mAcm-2}
\end{center}


\end{document}